\documentclass[a4paper,fleqn,usenatbib]{mnras}
\usepackage{newtxtext,newtxmath}
\usepackage[T1]{fontenc}
\usepackage{ae,aecompl}
\usepackage{graphicx}	
\usepackage{amsmath}	
\usepackage{amssymb}	
\usepackage{float}

\newcommand{\vs}{vs.}
\newcommand{\mass}{$M_*$}
\newcommand{\msun}{$M_{\sun}$}
\newcommand{\Sersic}{S\'ersic}
\newcommand{\Sig}{$\Sigma_1$}
\newcommand{\DelSig}{$\Delta\Sigma_1$}
\newcommand{\Delmue}{$\Delta\langle\mu_e\rangle$}
\defcitealias{Fisher2016}{FD16}
\defcitealias{Kormendy2016}{K16}
\defcitealias{Gadotti2009}{G09}

\title[Bulges and central density]{Structural and Stellar Population Properties vs. Bulge Types in Sloan Digital Sky Survey Central Galaxies}

\author[Y. Luo et al.]{Yifei Luo$^{1,2,3}$\thanks{E-mail: yifeiluo@ucsc.edu},
S. M. Faber$^{4}$,
Aldo Rodr\'iguez-Puebla$^{2,5}$,
Joanna Woo$^{6}$,
Yicheng Guo$^{7}$,
\newauthor
David C. Koo$^{4}$,
Joel R. Primack$^{8}$,
Zhu Chen$^{9}$,
Hassen M. Yesuf$^{4}$,
Lin Lin$^{10}$,
\newauthor
Guillermo Barro$^{11}$,
Jerome J. Fang$^{12}$,
Viraj Pandya$^{2}$,
M. Huertas-Company$^{13}$,
\newauthor
Shude Mao$^{14,1}$
\\
$^{1}$National Astronomical Observatories, Chinese Academy of Sciences, 20A Datun Road, Chaoyang District, Beijing 100012, China\\
$^{2}$Department of Astronomy and Astrophysics, University of California at Santa Cruz, Santa Cruz, CA 95064, USA\\
$^{3}$School of Astronomy and Space Science, Nanjing University, Nanjing, Jiangsu 210093, China\\
$^{4}$UCO/Lick Observatory, Department of Astronomy and Astrophysics, University of California, Santa Cruz, CA 95064, USA\\
$^{5}$Instituto de Astronom\'ia, Universidad Nacional Aut\'onoma de M\'exico, A. P. 70-264, 04510, M\'exico, D.F., M\'exico\\
$^{6}$Department of Physics and Astronomy, PO Box 1700 STN CSC, Victoria BC V8W 2Y2, Canada\\
$^{7}$Department of Physics and Astronomy, University of Missouri, Columbia, MO 65211, USA\\
$^{8}$Physics Department, University of California, Santa Cruz, CA 95064, USA\\
$^{9}$Shanghai Key Lab for Astrophysics, Shanghai Normal University, 100 Guilin Road, Shanghai 200234, China\\
$^{10}$Shanghai Astronomical Observatory, Chinese Academy of Sciences, Shanghai 200030, China\\
$^{11}$Department of Physics, University of the Pacific, 3601 Pacific Avenue, Stockton, CA 95211, USA\\
$^{12}$Orange Coast College, Costa Mesa, CA 92626, USA\\
$^{13}$GEPI, Observatoire de Paris, CNRS, Universit\'e Paris Diderot, 61, Avenue de l'Observatoire, F-75014, Paris, France\\
$^{14}$Department of Astronomy and Tsinghua Centre for Astrophysics, Tsinghua University, Beijing 100084, China}

\date{Accepted XXX. Received YYY; in original form ZZZ}

\pubyear{2019}

\begin{document}
\label{firstpage}
\pagerange{\pageref{firstpage}--\pageref{lastpage}}
\maketitle

\begin{abstract}

  This paper studies pseudo-bulges (P-bulges) and classical bulges (C-bulges) in Sloan Digital Sky Survey central galaxies using the new bulge indicator \DelSig, which measures relative central stellar-mass surface density within 1 kpc. We compare \DelSig\ to the established bulge-type indicator \Delmue\ from \citet[][]{Gadotti2009} and show that classifying by \DelSig\ agrees well with \Delmue.  \DelSig\ requires no bulge-disk decomposition and can be measured on SDSS images out to $z = 0.07$.  Bulge types using it are mapped onto twenty different structural and stellar-population properties for 12,000 SDSS central galaxies with masses 10.0 < log\ \mass/\msun\ < 10.4.  New trends emerge from this large sample.  Structural parameters show fairly linear log-log relations \vs\ \DelSig\ and \Delmue\ with only moderate scatter, while stellar-population parameters show a highly non-linear ``elbow" in which specific star-formation rate remains roughly flat with increasing central density and then falls rapidly at the elbow, where galaxies begin to quench.  P-bulges occupy the low-density end of the horizontal arm of the elbow and are universally star-forming, while C-bulges occupy the elbow and the vertical branch and exhibit a wide range of star-formation rates at fixed density.  The non-linear relation between central density and star-formation rate has been seen before, but this mapping onto bulge class is new.  The wide range of star-formation rates in C-bulges helps to explain why bulge classifications using different parameters have sometimes disagreed in the past.  The elbow-shaped relation between density and stellar indices suggests that central structure and stellar-populations evolve at different rates as galaxies begin to quench.

\end{abstract}


\section{Introduction}
\label{sec:1}
The centers of galaxies are uniquely interesting regions of the universe.  They are the bottoms of the deepest potential wells, they have the highest baryon densities on kiloparsec scales, they form stars under conditions that are very different from galactic disks, and they enable the growth of super-massive black holes.  For all these reasons, we would like to understand their properties in detail.

Important information has emerged about galaxy centers from two rather different directions. The first approach has focused on centers as distinct objects and has studied them using \emph{multiple} central properties such as morphologies, density profiles and other structural indices, stellar populations, dust and gas content, and star formation rates \citep[e.g.,][]{Kormendy2004, Athanassoula2005, Fisher2006, Fisher2008, Fisher2009, Fisher2010, Fisher2011, Fabricius2012}. A major result is that these disparate parameters are well enough correlated to justify a classification scheme that arranges galaxies into four bins -- no-bulge systems, pseudo-bulges, classical bulges, and ellipticals -- according to the prominence of a central, dynamically hot stellar population \citep{Kormendy2004}.   Recent reviews of the criteria for classifying galaxies on this system are given by \citet[][FD16]{Fisher2016} and by \citet[][K16]{Kormendy2016}. 

The second approach attempts to reduce bulge structure to a \emph{single number}, central stellar surface density within 1 kpc (\Sig). Previous work has shown two quite tight relations that separately describe star-forming galaxies and quenched galaxies in the \Sig-\mass plane  \citep[e.g.,][]{Cheung2012, Barro2013, Fang2013, vanDokkum2014, Tacchella2015, Woo2019, Mosleh2017, Whitaker2017, Lee2018}.  An advantage of \Sig\ is that it does not require high angular resolution or bulge-disk decomposition, which enables simple and robust measurements to be made out to $z = 0.07$ in SDSS \citep{Fang2013} and to $ z = 3$ in CANDELS \citep{Barro2017}, distances where the standard bulge-classification method cannot be used reliably with current data. In return for accepting less detailed information about the centers, the \Sig\ method has been able to assemble large, volume-limited data sets that better illuminate connections with global properties and evolutionary trends.   A major finding is that the two separate scaling laws that relate \Sig\ to galaxy mass \mass\ for quenched and star-forming galaxies have  remained very similar since $z = 3$ \citep{Barro2017}.   It thus appears that the manner in which galaxies build their central densities is an ancient process that has been in place for a long time.

The first purpose of this paper is to compare the  \Sig\ and pseudo-bulge/classical-bulge approaches.  Do they order galaxies by central properties in the same way?   A problem with the pseudo-bulge/classical-bulge method is that multiple criteria sometimes disagree.  Which criteria, if any, correlate best with  \Sig, and why?  Is  \Sig\ by itself a useful bulge classifier for SDSS when comparing it to SDSS indices that have not been used for bulge studies before? Finally, do recent insights on galaxy evolution from SDSS and other large surveys shed light on how bulges are evolving?

\begin{figure*}
    \includegraphics[width=1.75\columnwidth]{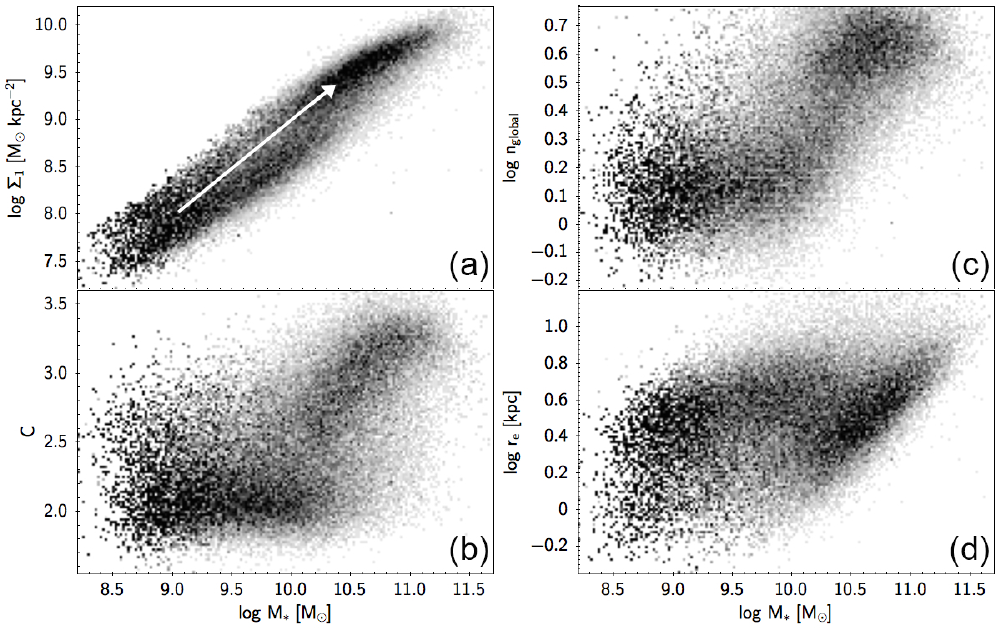}
    \caption{Panels a, b, c and d plot \Sig, concentration, global \Sersic\ index, and radius \vs\ stellar mass for face-on, central, non-interacting SDSS galaxies with $0.02 < z < 0.07$ as described in Table~\ref{table:sample} (49,567 objects).  The gray-scale represents the number density weighted by the completeness correction; all bulge types are included.  Two features are apparent.  Note the narrowness of the \Sig-\mass\ relation, as well as the two ``islands'' seen most strongly in \Sig, concentration, and global \Sersic\ index.  The upper ridgeline in panel a is populated mainly by quenched and quenching galaxies \citep{Fang2013, Barro2017}, but the islands in this figure are based on structure, not star-formation rate, and their membership is not exactly the same as the red-sequence/blue-cloud division based on star formation.  We have named the valley here the structural valley (SV) to distinguish it from the green valley of star formation.  The arrow represents a schematic evolutionary track in \Sig\ \vs\ \mass.}
    \label{fig:figure1}
\end{figure*}

Figure~\ref{fig:figure1} plots the \Sig-\mass\ relation and three other structural relations for a sample of nearby SDSS galaxies. The gray-scale represents the number density weighted by a completeness correction that is described below. Two features are apparent. One is the extreme narrowness of the \Sig-\mass\ relation.  Since galaxies grow significantly in mass, they must also increase in \Sig\ in order to stay on this relation; in other words, the \Sig-\mass\ relation for star-forming galaxies is an approximate evolutionary track \citep{Barro2017}.  The white arrow shows this schematically.  The second feature is two ``islands'', seen strongly in three of the panels and separated by a weak valley. For concentration and global \Sersic\ index, the upper  island is associated with quenched early-type galaxies while the lower island is associated with star-forming galaxies. It has been shown that the upper island/ridgeline in \Sig-\mass\ also consists mainly of quenched and near-quenched galaxies, while the lower \Sig\ island consists mainly of star-forming galaxies \citep[e.g.,][]{Cheung2012, Fang2013, vanDokkum2014, Barro2017}.  It is also known that star-formation rates divide massive galaxies into the red-sequence and blue-cloud, which are separated by the green valley (GV) \citep[e.g.,][]{Strateva2001, Blanton2003, Kauffmann2003a, Bell2004, Faber2007, Wyder2007, Bell2008, Brammer2009, Brammer2011}. The three valleys seen in Figure~\ref{fig:figure1}, however, are based on structure, not star-formation rates, and Section~\ref{sec:7.3} will show that the membership of galaxies in the \Sig\ islands is not exactly the same as membership in the red sequence and blue cloud.  More precisely, the islands contain all the red, quenched galaxies, but they contain additional star-forming galaxies as well.  A structural division similar to that shown here was seen in plots of concentration and \Sersic\ index \vs\ stellar population indices \citep{Kauffmann2003b,Baldry2006,Driver2006,Ball2008}, but the difference between it and the stellar-population green-valley division was not stressed.    We call this new feature the \emph{structural valley} (SV) to distinguish it from the green valley of star formation, and its relationship to pseudo-bulges and classical bulges is discussed in Section~\ref{sec:7.3}.

It would seem from Figure~\ref{fig:figure1}a that galaxies in general are evolving along the \Sig-\mass\ relation (see arrow), and therefore that \Sig\ must increase during a galaxy's lifetime. Furthermore, since the number of \emph{quenched} galaxies is also increasing with time \citep[e.g.,][]{Bell2004,Faber2007}, there must be a net flow of galaxies \emph{all along the track,} including from the main branch onto the upper island.  The remainder of this paper will show that the \Sig\ track encompasses all bulge types, from no-bulges and pseudo-bulges at the low-\Sig\ end to classical bulges and ellipticals on the island.  This implies that galaxies generally evolve smoothly in bulge type as well as \Sig, but previous literature on bulge types has not usually emphasized this possibility.  Rather, a frequent picture for the origin of bulge types envisions \emph{separate mechanisms} whereby pseudo-bulges (hereafter P-bulges) form from no-bulges (hereafter N-bulges) by internal secular evolution of galaxy discs whereas classical bulges (hereafter C-bulges) form via major mergers \citepalias{Fisher2016, Kormendy2016}, which not all galaxies would necessarily undergo.  Another proposal is that C-bulges form early via gas-rich instabilities and mergers \citep[e.g.,][]{Elmegreen2008, Dekel2009, Forbes2014, Ceverino2015} while P-bulges form more gradually over time, undisturbed, in less dense regions \citepalias{Fisher2016}.  If this were strictly true, C-bulges and P-bulges would be on separate evolutionary paths, and P-bulges might again not evolve into C-bulges.  Yet a third picture retains secular evolution and mergers as separate mechanisms but varies their importance smoothly with time and mass.  When galaxies are low-mass and gas-rich, it is said that mergers have limited ability to build stellar bulges \citep{Robertson2006}, and secular evolution is the major bulge-building mechanism.  As mass grows and gas content declines, mergers are increasingly able to puff disks into hot stellar systems, and bulges begin to grow mainly by mergers \citep{Hopkins2009a,Hopkins2009b}.  This version of the two-mechanism picture allows -- even requires -- that many P-bulges would evolve into C-bulges.  Indeed, \citet{Kormendy2010b} wondered whether mergers are so effective that \emph{all} massive galaxies should by now have developed classical bulges, contrary to what is seen.  

In contrast to the multiple bulge-building paths entertained in the bulge-classification literature, the \Sig\ literature has stressed an unbroken continuum in which \emph{all} galaxies are continuously building their centers in basically the same fashion, as envisioned in the bulge-evolution picture.   However, not much attention has been paid in this literature to the detailed mechanism(s) that do this. \citet{Fang2013} and \citetalias{Kormendy2016} have stressed that the rise of \Sig\ is a natural consequence of entropy growth in self-gravitating systems whereby a multitude of processes -- such as violent relaxation, disk instabilities, and gravitational interactions between bars, spirals arms, and stars, etc. -- permit sub-components to exchange energy and angular momentum among themselves, inevitably driving up central density while at the same time building an extended outer envelope.  Gaseous dissipation involves actual \emph{net energy loss} on top of this and compounds the concentration process still further, especially at early times when galaxies are gas-rich and dissipative compaction can occur \citep{Dekel2009, Zolotov2015, Barro2017}. Although there are mechanisms that can reduce \Sig, their effect is generally small, and they mostly occur after a galaxy is quenched \citep{vanDokkum2015, Barro2017}. Thus, according to this picture, \Sig\ acts as a kind of clock for galaxy evolution: it can easily go up, but it can never go down (by very much).

The previous discussion has glossed over an important point, namely, that P-bulges are actually of two types: boxy/peanut bulges, which form via vertical dynamical instabilities in a bar, and disky bulges, which form from gas brought to the central regions via secular evolution \citep{Kormendy2004,Athanassoula2005}.  These two types of bulges may affect different bulge indicators differently, and we return to the question of barred \vs\ unbarred galaxies in Section \ref{sec:4}.

We further note that even though galaxies may be evolving through the various bulge types, not all massive galaxies have evolved all the way to the C-bulge or elliptical stage.  Many nearby massive galaxies have P-bulges, e.g., NGC 1097 and M51  \citep{Kormendy2010a,Kormendy2010b, Kim2014, Salo2015, Gadotti2019, Querejeta2019}; see also Figure \ref{fig:figure16} below.  Why some massive galaxies may have evolved to earlier bulge types but others have not is an open question; the data from this study may help to answer it.

Since valuable information about galaxy centers has come from both the \Sig\ and bulge-classification approaches, the first goal of this paper is to see whether the two approaches agree. To accomplish this, a sample of objects is needed that has been classified both ways. Fortunately, such a sample is available from the work of \citet[][G09]{Gadotti2009}, who measured bulge types for nearly 1000 SDSS galaxies based on \Delmue, one of the classic parameters used to classify galaxies in the bulge-classification method (see Section~\ref{sec:2} for the definition of \Delmue).  Section~\ref{sec:4} compares \Delmue\ to \DelSig\ (which is \Sig\ with mass-trend removed), and it appears that both parameters measure approximately the same thing, namely, central stellar mass density. 

Having established that \DelSig\ is a valid indicator for SDSS bulges, we then move on to the second part of the paper, which maps 20 different SDSS properties onto bulge types for $\sim$12,000 SDSS galaxies.  New trends emerge from this large and homogeneous sample.  The main result is that structural parameters are generally found to correlate closely with \DelSig, which is itself a structural parameter, but that spectral indices of C-bulges are bifurcated, with some C-bulges being quenched and others (very) actively star-forming. This is consistent with previous studies of star-formation rate as a function of \DelSig\ and helps to explain why classification of C-bulges has proved difficult -- they are not a homogenous class.  The bottom line is that the structure and stellar populations of bulges are highly correlated, but not in a linear fashion.

This paper is organized as follows. The data and sample selection are described in Section~\ref{sec:2}. Section~\ref{sec:3} uses the structural valley to determine the trend of \Sig\ \vs\ mass and defines a residual \DelSig\ relative to the valley.  \DelSig\ is our preferred parameter to characterize the structural state of galaxy bulges.  It is compared to the traditional parameter \Delmue\ in Section~\ref{sec:4} using the G09 sample, and agreement is generally good.  Further comparisons in Section~\ref{sec:5} and Section~\ref{sec:6} reinforce this using 20 structural and stellar-population parameters from SDSS and other data.  Finally Section~\ref{sec:5} and Section~\ref{sec:6} show trends for the whole SDSS sample in the mass range 10 < log\ \mass/\msun\ < 10.4.    Implications are discussed in Section~\ref{sec:7}, and a summary is given in Section~\ref{sec:8}. Unless otherwise noted, all structural measurements are based on the SDSS $i$-band. We adopt a concordance $\Lambda$CDM cosmology: $H_0$ = 70 km $\rm s^{-1}\ Mpc^{-1}$, $\Omega_m$ = 0.3 and $\Omega_\Lambda$ = 0.7.
    
\section{data and sample section}
\label{sec:2}
Two main samples of galaxies are used in this paper.  The first is a large SDSS sample used to study the general properties of galaxies \vs\ central density.  It is selected from the SDSS DR7 catalog \citep{Abazajian2009}. To avoid seeing degradation, we limit the redshift range to $0.02 < z < 0.07$ \citep{Fang2013}. Besides the redshift cut, galaxies are limited to log\ \mass/\msun\ > 9.5, axis-ratio \textit{b/a} > 0.5, and single-\Sersic\ index 0.5 < \textit{n} < 6. The decision was also made to focus on \emph{central} SDSS galaxies for this first analysis.  It is known that environment affects both stellar populations \citep[e.g.,][]{Blanton2009,Woo2015} and structural properties \citep{Woo2017}, but the detailed mechanisms are not yet known. We therefore prefer to use the simpler life histories of field galaxies to formulate a basic picture of bulge properties, against which satellite galaxies can be compared in future work.   Satellite galaxies are accordingly excluded based on the group designation ($M_{\mathrm{rank}} > 1$) according to \citet{Yang2012}. Merging galaxies are also excluded for similar reasons, using the classification ($P_{\mathrm{MG}} < 0.1$) from Galaxy Zoo \citep[][]{Lintott2011}. After applying these criteria, 41,272 objects are left in this sample.

\begin{figure}
    \includegraphics[width=\columnwidth]{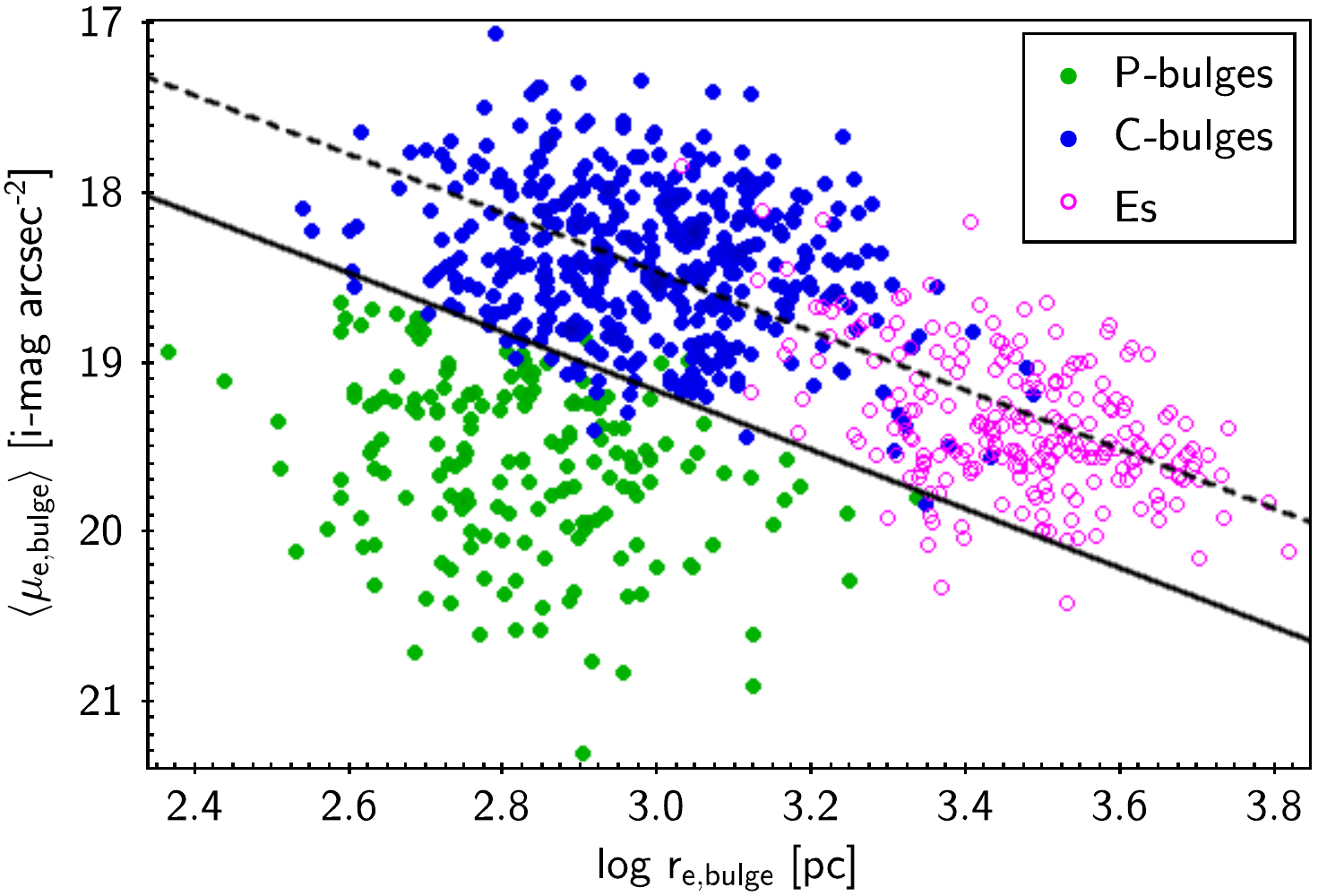}
    \caption{Mean bulge effective surface brightness \vs\ bulge effective radius (Kormendy relation, \citealt{Kormendy1977}) using data for SDSS galaxies derived from the catalog of \citet[][G09]{Gadotti2009}. Green, blue, and  magenta points indicate P-bulges, C-bulges, and ellipticals as classified by \citetalias{Gadotti2009}. \Delmue\ is the residual value of $\langle\mu_e\rangle$ relative to the solid line (which is taken from \citetalias{Gadotti2009}). The dashed line above it schematically represents the ridgeline of galaxies identified by \citet{Kormendy1977} as consisting of classical bulges and elliptical galaxies.
        }
    \label{fig:figure2}
\end{figure}

\begin{figure}
    \includegraphics[width=\columnwidth]{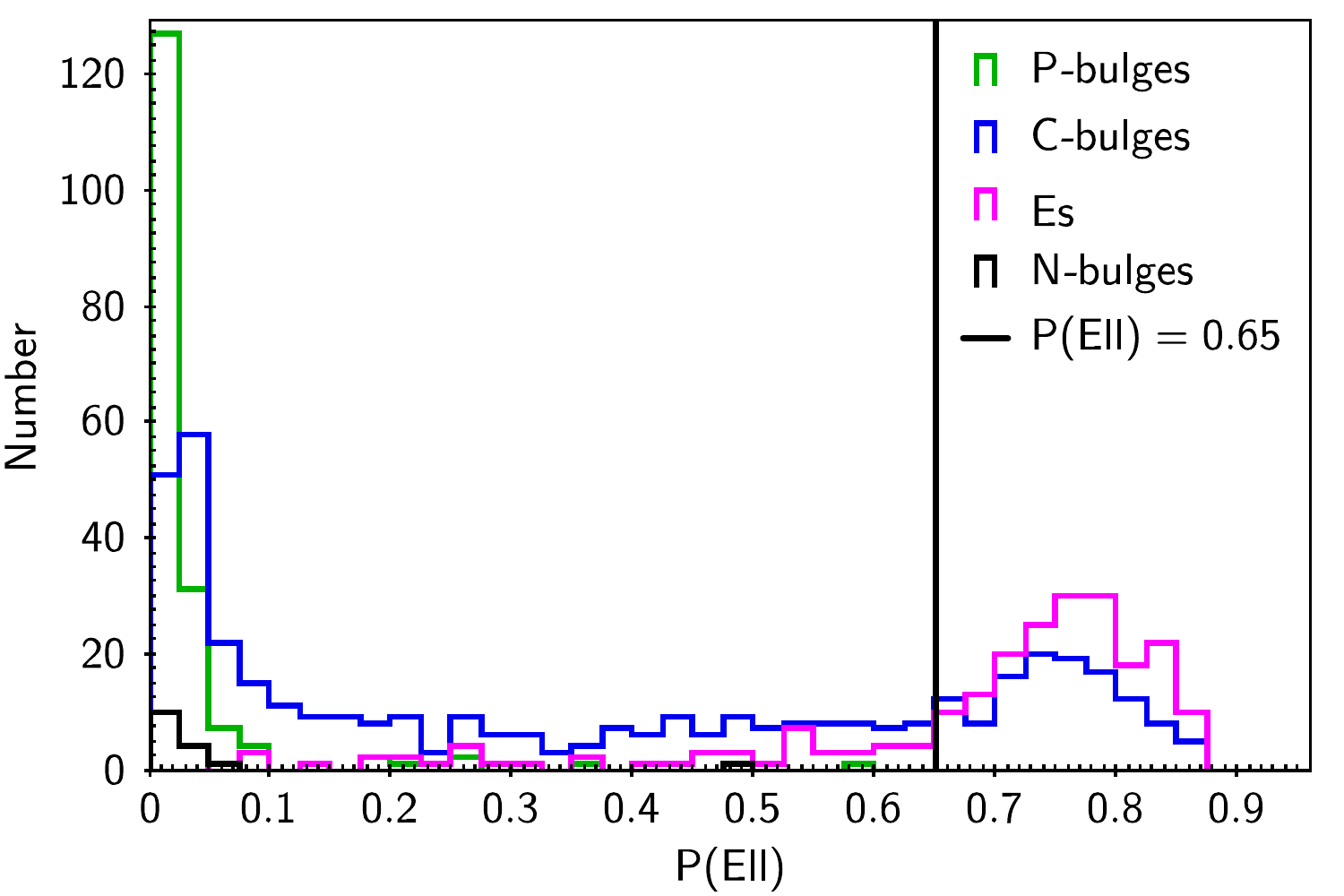}
    \caption{The histogram of the probability of a galaxy's being an elliptical, P(Ell), for G09 galaxies. Probabilities are taken from the morphological study of \citet{Huertas-Company2011}.  Black, green, blue and magenta histograms represent N-bulges, P-bulges, C-bulges, and E's, respectively, based on the G09 classifications. N-bulges and P-bulges are cleanly distinguished from E's, but C-bulges (blue line) are a mixture of types.  We could not find any other measurable parameter to differentiate E-like C-bulges from true E's in the G09 sample, and so all SDSS galaxies with P(Ell) > 0.65 (vertical line) are classed as E's.  N-bulges are classified by the definition of $B/T$ = 0 in \citetalias{Gadotti2009}, which makes them rare in the G09 mass range. }
    \label{fig:figure3}
\end{figure}

Integrated photometry (model magnitude) and surface brightness profiles in \textit{ugriz} bands for SDSS galaxies are obtained from the SDSS database and corrected for Galactic extinction. \textit{K}-corrections are applied using the \textit{k}-correction code v4.2 from \citet{Blanton2007}. We compute the cumulative light profile and interpolate smoothly in order to calculate the total light within 1 kpc. The stellar mass surface density within 1 kpc, \Sig\footnote{The \Sig\ value for SDSS DR7 galaxies is available online as supplementary material.}, is computed using the total light within 1 kpc in $i$-band and \textit{M/$L_i$} from \citet{Fang2013}. We also add $NUV$ magnitude for those galaxies which have observations in the \textit{GALEX} database and apply the Galactic correction and \textit{K}-correction as for the SDSS photometry. Spectroscopic data including redshifts, stellar masses, fiber velocity dispersions, D$_n$4000, H$\delta_A$, H$\alpha$ equivalent width, [OIII], H$\beta$, [NII], H$\alpha$ fluxes, global SSFR and fiber SSFR are obtained from the MPA/JHU DR7 value-added catalog\footnote{\url{http://www.mpa-garching.mpg.de/SDSS/DR7/}}.  We exclude galaxies whose spectra with median per-pixel S/N < 10 when using spectroscopic parameters. Structural parameters in the $i$-band, such as concentration index (defined as $R90/R50$), global \Sersic\ index and effective radius of the galaxies in kpc (converted from the single \Sersic\ fit), are taken from the NYU Value-Added Galaxy Catalog (NYU-VAGC) DR7\footnote{\url{http://sdss.physics.nyu.edu/vagc/}} \citep[][]{Blanton2005, Adelman-McCarthy2008, Padmanabhan2008}. We also add the bulge-disk decompositions and $b/a$ values from \citet{Simard2011}. Internal reddening corrections from \citet{Oh2011} are applied to all colors but not to emission lines or to absorption indices. 

\begin{figure}
    \includegraphics[width=\columnwidth]{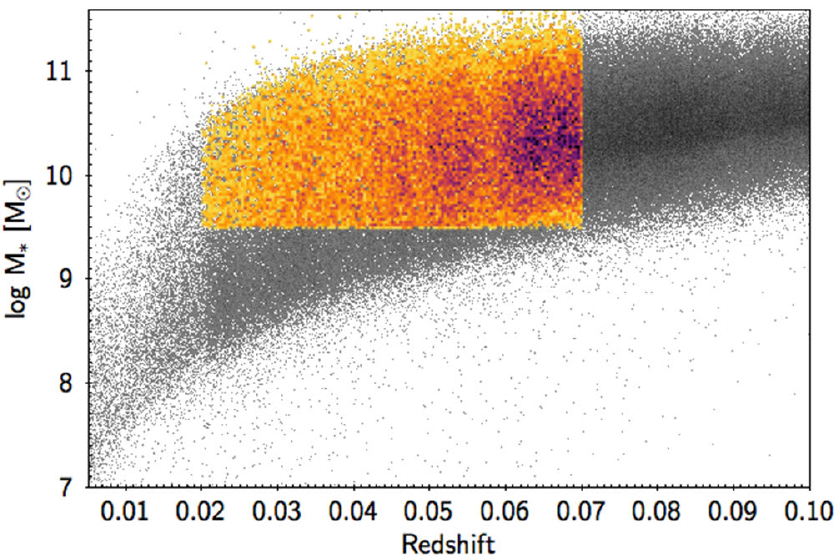}
    \caption{Stellar mass \vs\ redshift, color-coded by raw number density. The grey background is all SDSS galaxies with $z < 0.1$ in all mass ranges, with upper and lower boundaries set by $14 < r < 17.5$. The yellow foreground is our final SDSS sample (Table~\ref{table:sample}) with $0.02 < z < 0.07$, \mass\ > $10^{9.5} M_{\sun}$, $b/a > 0.5$, centrals only, and good-quality single-\Sersic\ fits. The dark band marks quenched galaxies, which are not detected as deeply as star-forming galaxies on account of their red colors; some low-mass objects of this type are missed at high redshift in the lower-right corner of the yellow region.  Conversely, the magnitude cutoff at upper left discriminates against high-mass objects at low redshift.  Significant volume-limited corrections are needed both below $10^{10} M_{\sun}$ and above $10^{11} M_{\sun}$ to correct for these effects (see text).}
    \label{fig:figure4}
\end{figure} 

\begin{table}
 \caption{Sample descriptions and sizes.}
 \label{table:sample}
 \begin{tabular}{|p{3.6cm}|p{3.1cm}|p{0.8cm}|}
  \hline
  Desciption & Criterion & N \\
  \hline
  SDSS sample$^{\textrm{a}}$ & &  \\
  Redshift limit & $0.02 \leq z \leq 0.07$ & 156,634 \\
  Magnitude limit & $14 \leq r \leq 17.5$ & 136,271\\
  Face-on galaxies & $b/a\geq0.5$ & 84,656 \\
  Good \Sersic\ fit & $0.5 < \textit{n} < 5.9$ & 80,098\\
  Central galaxies & $M_{\mathrm{rank}}=1$ & 53,519\\
  Non-interacting galaxies &  $P_{\mathrm{MG}}<0.1$ & 49,567 \\
  SDSS sample & log\ \mass/\msun\ > 9.5 & 41,272\\
  Mass limited SDSS sample$^{\textrm{b}}$ & 10 < log\ \mass/\msun\ < 10.4 & 12,421\\
    & & \\
  G09 sample$^{\textrm{c}}$ & Galaxies from \citetalias{Gadotti2009} & 860\\
  \hline
  \end{tabular}
\small $^{\textrm{a}}$Cross-matched with all mentioned catalogs.\\
\small $^{\textrm{b}}$About 1000 galaxies with median per-pixel S/N < 10 in this sample are excluded when using spectroscopic parameters.\\
\small $^{\textrm{c}}$Cross-matched with all mentioned catalogs but \textit{without} the criteria listed above.
\end{table}
   
Since we aim to study correlations between \Sig\ and bulge type, we also need a separate sample with pre-classified bulge types for reference. \citetalias{Gadotti2009} provides a suitable reference sample of nearly 1000 SDSS galaxies that have been classified into four bulge types (N-bulges, P-bulges, C-bulges and E's, i.e., pure bulges) based on careful bulge-disk-bar decompositions. E's and N-bulges are identified by \citetalias{Gadotti2009} according to whether a bulge or disk solely is needed in the decomposition. If an accurate decomposition does not require a bulge it is termed a bulgeless galaxy (N-bulge). If it does not require a disk it is an elliptical. The remaining objects need both bulge and disk.  Among these, P-bulges are distinguished from C-bulges using \Delmue\ from the Kormendy relation \citep{Kormendy1977}, as shown in Figure~\ref{fig:figure2}:
\begin{equation}
\label{equ:delmue}
\Delta\langle\mu_e\rangle=\langle\mu_e\rangle-1.74 {\rm log} r_e-13.95,
\end{equation}
where $\langle\mu_e\rangle$ and \Delmue\ are in $i$-magnitude arcsec$^{-2}$,  $r_e$ is in parsec, and the equation of the line is from \citetalias{Gadotti2009}.\footnote{Note that \citetalias{Gadotti2009} does not provide the mean effective surface brightness within the bulge effective radius $\langle\mu_e\rangle$ in the on-line catalog, but rather effective surface brightness at the bulge effective radius $\mu_e$ only. Thus, we compute $\langle\mu_e\rangle$ ourselves using $\mu_e$ and bulge \Sersic\ index from the on-line catalog.} Traditionally, a combination of both structural and spectral criteria has been used to distinguish P-bulges from C-bulges \citepalias{Fisher2016,Kormendy2016}. The G09 types in contrast are based purely on a structural criterion, i.e., \Delmue.  We show later (Section~\ref{sec:6}) that structural criteria and stellar-population criteria sometimes disagree systematically in classifying bulges.  It should therefore be kept in mind when assessing the agreement between our classifications and other parameters that our reference sample is biased toward using a structural criterion.

The bulge-disk-bar decomposition in \citetalias{Gadotti2009} also included bars as separate structures. We elect to leave out the bar-component contribution to central density when computing \DelSig\ for G09 galaxies, but we verify in  Figure~\ref{fig:figure9} below that this choice has not moved barred galaxies systematically off the C-bulge correlation. In the interest of maximizing the number of galaxies, we have likewise retained both satellites as well as centrals in the G09 sample. 

We preserve the bulge types from \citetalias{Gadotti2009} when studying that sample, and this includes using his E classification. When studying the SDSS sample, it is desirable to retain the E/C-bulge distinction, but high-quality bulge-disk decompositions for SDSS are not available.  Therefore we use the probability of a galaxy's being an elliptical, P(Ell),  from \citet{Huertas-Company2011}. Figure~\ref{fig:figure3} shows the P(Ell) histogram for the G09 sample as a function of G09 bulge type. N-bulges and P-bulges are well separated from Es, but a large population of C-bulges overlaps with E's, as shown in the figure. We tested the G09 sample extensively to see if any other \textit{structural} parameter, such as SDSS concentration, $B/T$, global \Sersic\ index, velocity dispersion, or effective mass surface density, could be used to distinguish E-like C-bulges from true E's, but found none that worked. We therefore elect to classify all SDSS galaxies with P(Ell) > 0.65 as E's. N-bulges are rare in the G09 sample due to the lower mass limit log\ \mass/\msun\ > 10. The plot of bulge frequencies in \citet[][]{Fisher2011} also indicates that N-bulges are rare when log\ \mass/\msun\ > 10. Thus, we do not try to make the N-bulge/P-bulge division in this paper but rather simply group the G09 N-bulges in with the P-bulges. 

The grey background in Figure~\ref{fig:figure4} shows the raw density distribution of SDSS galaxies with $z$ < 0.1 in all mass ranges. The  foreground plot in yellow shows the density plot for the final sample with all cuts applied.  Significant volume-limited corrections are needed for both high-mass and low-mass galaxies (see caption).   We therefore calculate \textit{$V_{\rm max}$} and \textit{$V_{\rm min}$} for each object using its redshift and the SDSS survey limits 14 < \textit{r} < 17.5, where \textit{$V_{\rm max}$} and \textit{$V_{\rm min}$} represent the maximum and minimum volume over which the object would be included in the survey. Each galaxy is assigned a completeness correction for both the bright and faint limit based on \textit{$V_{\rm max}$} and \textit{$V_{\rm min}$}. SDSS galaxies in all density plots in this paper are weighted by this correction. In much of what follows, a mass-limited SDSS sub-sample is selected with 10.0 < log\ \mass/\msun\ < 10.4 to highlight masses where the structural bimodality is most prominent (cf.~Figure~\ref{fig:figure1}).  This mass range is approximately 67$\%$ complete in the range 10.0 < log\ \mass/\msun\ < 10.1, 80$\%$ complete in the range 10.1 < log\ \mass/\msun\ < 10.2 and 100$\%$ complete in the range 10.2 < log\ \mass/\msun\ < 10.4. The selection criteria and resulting sample sizes are summarized in Table~\ref{table:sample}.

To summarize, there are two main samples used in the remainder of this paper to compare bulge types to structural and spectral indices.  One is the small G09 sample with high-quality bulge-disk decompositions.  This sample of about 1,000 galaxies covers all  stellar masses above $10^{10}$ \msun\ and contains both central and satellite galaxies.  The second is the much larger mass-limited SDSS sample in the range $10^{10.0-10.4}$ \msun, which lacks accurate bulge-disk decompositions and consists of centrals only.  

\section{The structural valley in the \Sig-\mass\ diagram}
\label{sec:3}
\begin{figure}
    \centering
    \includegraphics[width=\columnwidth]{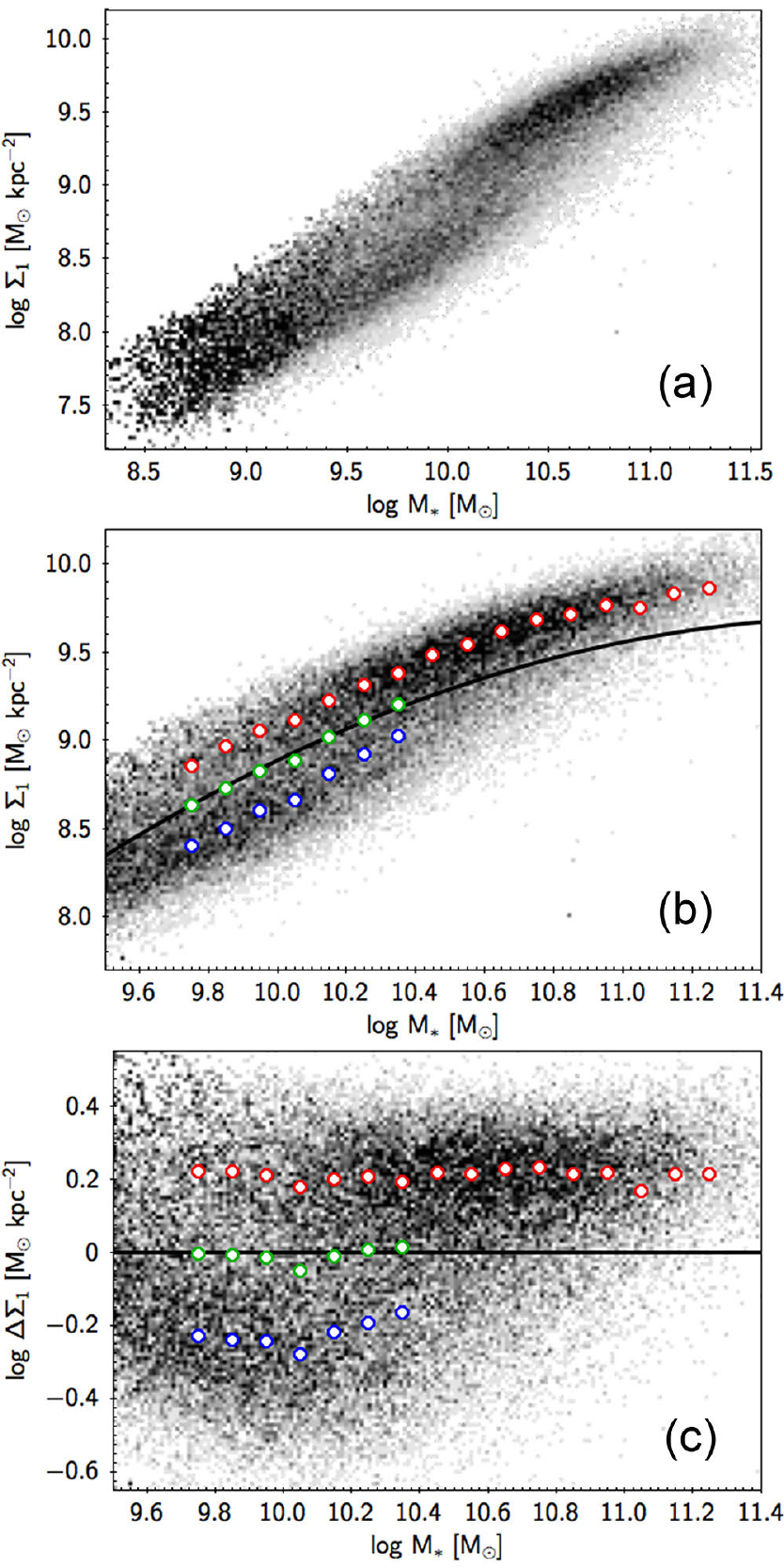}
    \caption{Panel a: \Sig\ \vs\ stellar mass for all SDSS galaxies, ellipticals included, repeated for reference from Figure~\ref{fig:figure1}a. Panel b: \Sig\ \vs\ stellar mass for the final SDSS sample with log\ \mass/\msun\ > 9.5. Panel c: \DelSig\ \vs\ stellar mass for the SDSS galaxies as in panel b. Points in all three panels are color-coded by number density weighted by the completeness correction. The red and blue points with white centers are the locations of the high-\Sig\ and low-\Sig\ populations determined by double Gaussian fitting (see Figure~\ref{fig:figure6}). The location of the structural valley (SV) is indicated by the green points with white centers, which are half-way between the red and blue points. The solid black line in panel b is a parabola  fitted to the ridgeline of the red points and shifted downward by 0.21 dex to match the SV.  Its equation is in the text. \DelSig\ is defined to be zero along this line, which flattens out the distributions \vs\ mass.}
    \label{fig:figure5}
\end{figure}

When using \Sig\ as a central density indicator for a wide range in stellar mass, it is important to remove any trend \vs\ mass first. \citet{Fang2013} showed that the \Sig\ ridgeline is tilted and that the threshold for quenching grows with stellar mass. \citet{Barro2017}, \citet{Tacchella2015} and \citet{Lee2018} confirmed this out to $z \sim$ 3. All four studies fitted the ridgeline with a simple power law in log-log coordinates. With more careful inspection, we observe a slight curvature in Figure~\ref{fig:figure1}a (repeated in Figure~\ref{fig:figure5}b).

\begin{figure}
    \includegraphics[width=\columnwidth]{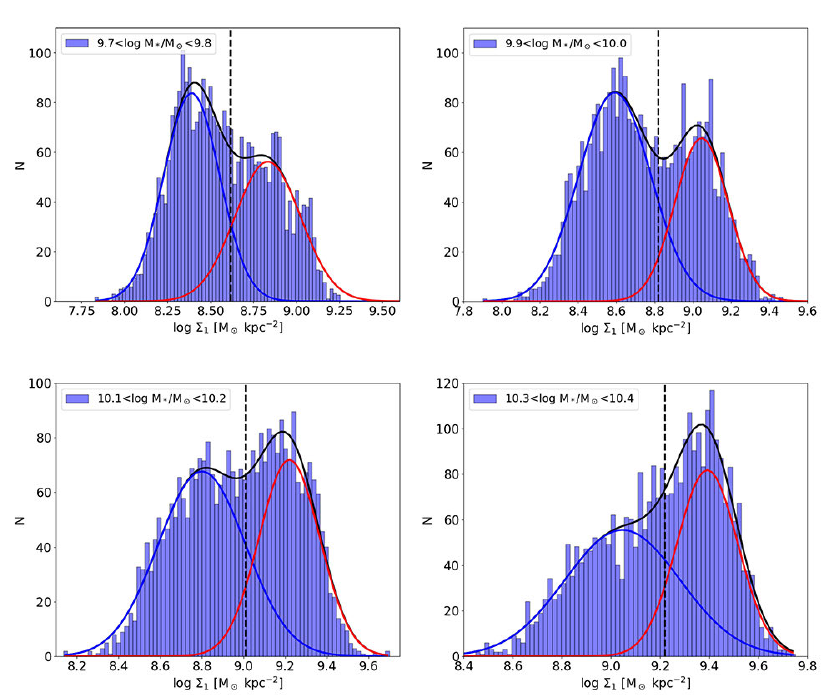}
    \caption{Histograms of \Sig\ with fitted double Gaussians in four illustrative mass bins of SDSS: 9.7 < log\ \mass/\msun\ < 9.8, 9.9 < log\ \mass/\msun\ < 10.0, 10.1 < log\ \mass/\msun\ < 10.2 and 10.3 < log\ \mass/\msun\ < 10.4. Completeness corrections have been applied, and only bulges are counted (ellipticals are excluded). Red and blue lines are the individual Gaussians; black lines are the sum. The black dashed lines show the valley, which is defined by the half-way point between the Gaussian centers. }
    \label{fig:figure6}
\end{figure}

A feature noted in Figure~\ref{fig:figure1} is a weak valley that separates quenched from star-forming galaxies. Figure~\ref{fig:figure5}a repeats Figure~\ref{fig:figure1} for reference, and Figure~\ref{fig:figure5}b is magnified to show more detail at the high-mass end. Figure~\ref{fig:figure6} shows histograms of vertical slices through the population at representative masses (ellipticals have been deleted from these slices in order to better equalize the height of the two peaks), which show that the maximum depth is only about 15$\%$.  As noted, in order to distinguish between the GV found in the color-magnitude diagram and the new valley found in \Sig-\mass, we have dubbed this feature the \textit{structural valley (SV)} since \Sig\ is a structural parameter, not a color.  The general trend, noted in the Introduction, of galaxies to evolve from star-forming to quenched \citep[e.g.,][]{Bell2004,Faber2007} then implies a net flow of galaxies evolving across the SV.

To remove the mass trend, we fit double Gaussians to the slices shown in Figure~\ref{fig:figure6}.  The SDSS sample in Table~\ref{table:sample} is used, E's are excluded, and the volume-limited density correction is applied.   Figure~\ref{fig:figure6} shows the fitting results for four illustrative mass slices: 9.7 < log\ \mass/\msun\ < 9.8, 9.9 < log\ \mass/\msun\ < 10.0, 10.1 < log\ \mass/\msun\ < 10.2 and 10.3 < log\ \mass/\msun\ < 10.4.  A single Gaussian is used when log\ \mass/\msun\ > 10.4 since the low-\Sig\ galaxies are too few in these mass slices. The two populations are not strongly separated, but the peaks (and hence the SV) appear to be well located. The location of the SV at each mass is defined as the half-way point between the two Gaussian centers.  This definition is adopted since it is more robust than the minimum itself to changes in the relative strengths of the peaks.

The peaks are shown as blue and red points in Figure~\ref{fig:figure5}b, and the SV is delineated by the green points half-way between them. For the first seven mass slices, for which both red points and blue points exist, we fit two linear parallel lines to red and blue points and find a 0.42 dex distance between them. A second-order polynomial is then used to fit all the red points, which represents the \Sig-\mass\ relation for the high-\Sig\ population. The location of the SV is taken relative to the high-\Sig\ ridge line defined by bulges only but shifted downward by 0.21 dex.  \DelSig\ is defined as the residual value of \Sig\ relative to the black line:  
\begin{equation}
\label{equ:delsig1}
{\rm log}\Delta\Sigma_1={\rm log}\Sigma_1+0.275({\rm log}M_*)^{2}-6.445{\rm log}M_*+28.059,
\end{equation}
with \Sig\ and \DelSig\ in $M_{\sun}$kpc$^{-2}$ and \mass\ in \msun.

Figure~\ref{fig:figure5}c shows the density plot of \DelSig\ \vs\ \mass. All trends are now flat, and \DelSig\ = 0 divides galaxies into the two structural clouds. There is a slight turn-up in the ridgelines at low mass, but \Sig\ will be used only above log \mass\ = 9.8, where the relations are quite flat. We will test \DelSig\ as a new parameter to differentiate P-bulges from C-bulges.  In our picture in which galaxies are moving in \Sig\ and \mass\ from lower left to upper right in Figure~\ref{fig:figure5}a,  \citep[][]{Fang2013, Barro2017}, \DelSig\ is an evolving quantity that measures the distance of a galaxy from the final quenched ridgeline.

\section{Comparing \Delmue\ and \DelSig}
\label{sec:4}

The next three sections of the paper compare the new parameter \DelSig\ to other variables, starting in this section with \Delmue\ from \citetalias{Gadotti2009}, which was defined in Figure~\ref{fig:figure2}.  Recall that the G09 sample was selected to provide a calibration sample of SDSS galaxies with known bulge types.  Therefore, we hope to find good agreement between \DelSig\ and \Delmue\ in this comparison.

\begin{figure}
    \includegraphics[width=\columnwidth]{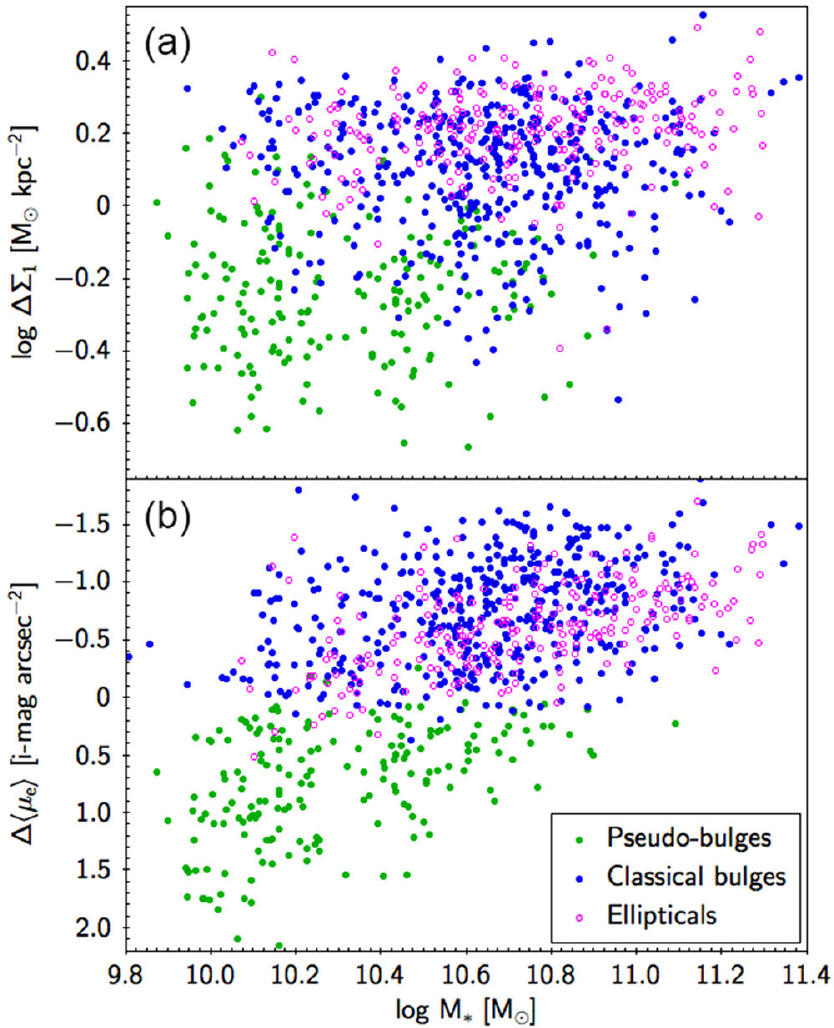}
    \caption{Panel a: \DelSig\ \vs\ \mass\ for G09 galaxies. Panel b: \Delmue\ \vs\ \mass\ for G09 galaxies. Points are color-coded according to the bulge classification in \citetalias{Gadotti2009}. Green and blue points represent P-bulges and C-bulges, and magenta circles are Es. The quenched ridgeline and the structural valley (SV) are more prominent in panel a, and E's are also located closer to the high-\DelSig\ ridgeline. Overall, however, the two parameters appear to be measuring roughly the same thing, especially if the residual mass trend in \Delmue\ were removed.
    }
    \label{fig:figure7}
\end{figure}

\begin{figure*}
    \includegraphics[width=2\columnwidth]{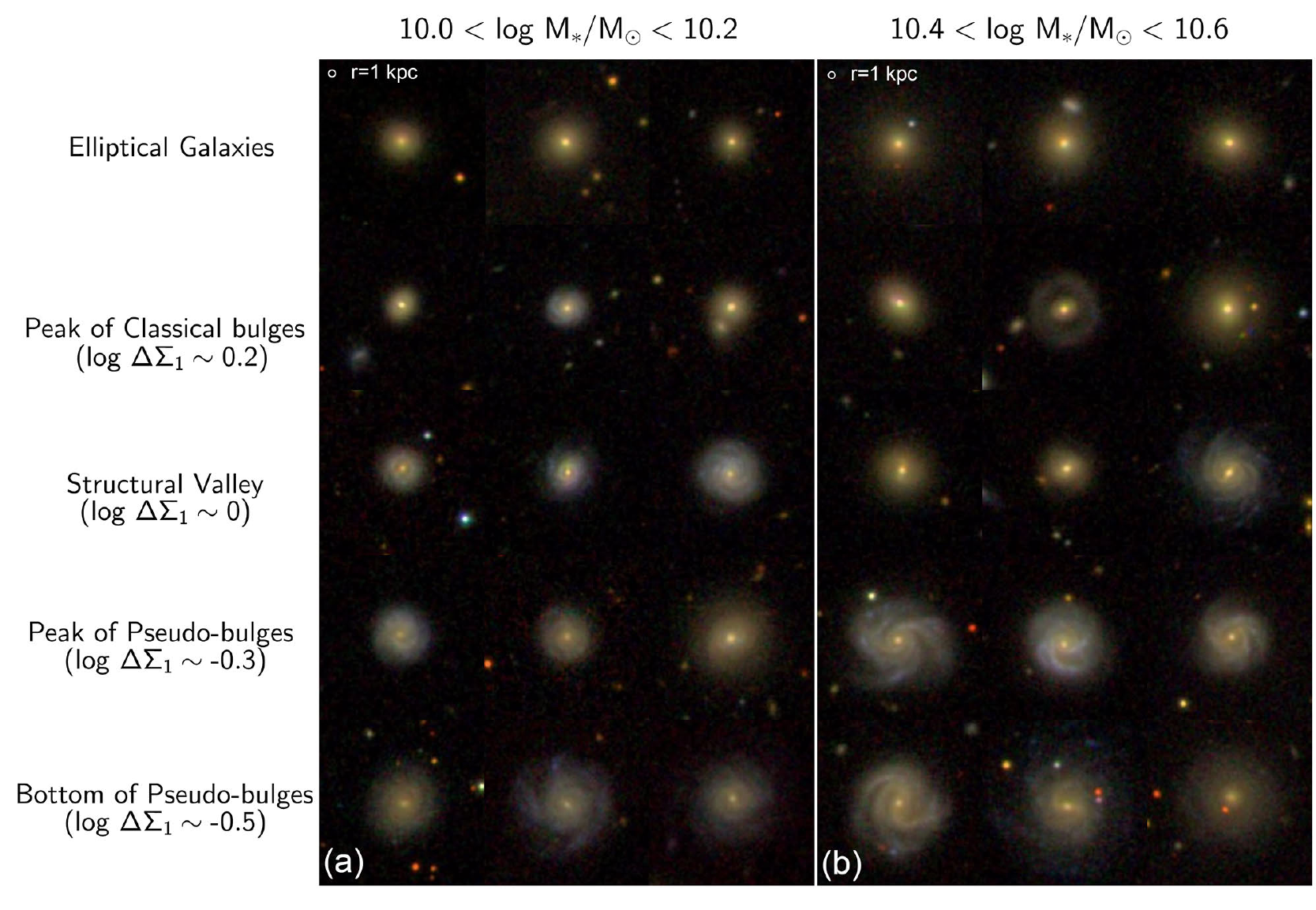}
    \caption{Sample SDSS postage stamps for G09 galaxies with redshifts $z \sim$ 0.04. Panel a shows galaxies with 10.0 < log\ \mass/\msun\ < 10.2, and panel b shows galaxies with 10.4 < log\ \mass/\msun\ < 10.6. The top row shows E's according to \citetalias{Gadotti2009} that also have P(Ell) > 0.65 according to \citet[][]{Huertas-Company2011}. The other four rows are arranged downward by  \DelSig. E's and C-bulges peak around log \DelSig\ $\sim$ +0.2, while  P-bulges peak around log \DelSig\ $\sim$ -0.3. Galaxies with log \DelSig\  $\sim$ 0 are in the structural valley (SV). Galaxies with log \DelSig\ $\sim$ -0.5 are at the bottom of the log \DelSig-\mass\ diagram. The two white circles in the upper left corners have a radius of 1 kpc, which is the region where \Sig\ is calculated.  Similar figures were presented in \citet{Fang2013} and \citetalias{Gadotti2009}.}
    \label{fig:figure8}
\end{figure*}

Figure~\ref{fig:figure7} compares \DelSig\ and \Delmue\ \vs\ stellar mass for the G09 galaxies.  Points are color-coded according to the bulge classification in \citetalias{Gadotti2009}. The agreement with the colored points is essentially perfect in the lower panel, as expected since the \citetalias{Gadotti2009} bulge types are based on this parameter. However, the quenched ridgeline and SV are more clearly defined in \DelSig-\mass. The C-bulge ridge is offset above theusing \Delmue\ but superimposes closely using \DelSig.  This latter agrees better with the claim by \citetalias{Kormendy2016} that the bulges of strong C-bulges are identical to E's (but see \citetalias{Gadotti2009} for a different view). \Delmue\ varies significantly with stellar mass, while \DelSig\ is more constant, the mass trend having been removed. This contributes to the fact that \Delmue\ tends to classify more galaxies as C-bulges at log\ \mass/\msun\ > 10.5 and fewer galaxies as C-bulges at log\ \mass/\msun\ < 10.5 than \DelSig. On the whole, however, the two parameters appear to be measuring the same thing, namely, central density.

Figure~\ref{fig:figure8} shows examples of SDSS postage stamps for G09 galaxies with redshifts $z \sim$ 0.04 for two mass bins 10.0 < log\ \mass/\msun\ < 10.2 and 10.4 < log\ \mass/\msun\ < 10.6. Galaxies from top to bottom are arranged by their \DelSig\ values. The two white circles in the upper left corner of the panels show the region where \Sig\ is calculated. The smooth progression of morphology from bottom to top row indicates that \DelSig\ sorts galaxies well by bulge prominence, which is as expected since Figure 9 in \citet{Fang2013} has shown a similar morphology transition using \Sig. Ordering galaxies by \Delmue\ gives substantially the same results \citepalias{Gadotti2009}, further confirming the similarity of  \DelSig\ and \Delmue. 

\begin{figure}
    \includegraphics[width=\columnwidth]{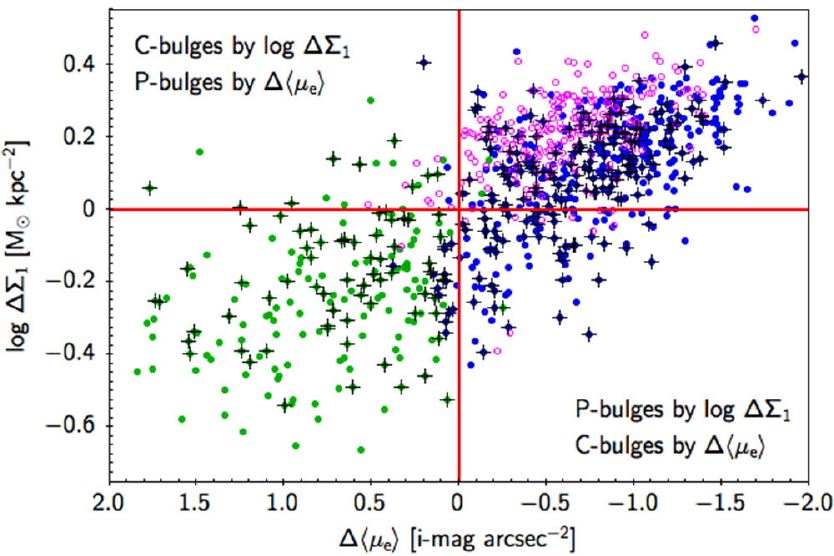}
    \caption{\DelSig\ is compared directly to \Delmue\ for the G09 galaxies. Points are color-coded according to the galaxy classifications in \citetalias{Gadotti2009}. Green points are P-bulges, blue points are C-bulges, and magenta circles are E's.  The reasonable agreement between \DelSig\ and \Delmue\ suggests that \DelSig\ can substitute for \Delmue\ as a bulge classification parameter. Some disagreement occurs in the upper left and lower right quadrants, where types disagree. Separate tests show that the residuals correlate with galaxy mass and radius, reflecting the fact that the mass trend has been removed from \DelSig\ but not from \Delmue\ (cf. Figure~\ref{fig:figure7}). Black crosses show barred galaxies according to \citetalias{Gadotti2009}. Their location relative to unbarred galaxies is discussed in the text.}
    \label{fig:figure9}
\end{figure}

Finally, Figure~\ref{fig:figure9} plots \DelSig\ \vs\ \Delmue\ directly for the G09 galaxies.  \DelSig\ follows \Delmue\ fairly closely but with some scatter.  Separate tests show that the residuals in this relation correlate mainly with galaxy radius and mass: larger and more massive galaxies lie to the lower right. These systematic residuals are due in part to the fact that mass/size trends have been removed from \DelSig\ but not from \Delmue\ (see discussion, Figure~\ref{fig:figure11}), and if that were done, the agreement in Figure~\ref{fig:figure9} would tighten. In other words, the suspicion is that, by applying information from additional parameters, it should be possible to predict \Delmue\ accurately from \DelSig.  This is confirmed by the work of \citet{Yesuf2019}, who use the random forest classifier to predict G09 bulge types from SDSS data, achieving an accuracy of 95\%.  As expected, \DelSig\ has the highest feature importance, followed by concentration and fiber velocity dispersion.

Barred galaxies identified by \citetalias{Gadotti2009} are indicated by the black crosses in Figure~\ref{fig:figure9}. \citetalias{Gadotti2009} models bars and bulges separately, but the \Delmue\ parameter incorporates only the mass in bulges whereas \DelSig\ uses the entire central mass.  It is therefore important to check whether there is an offset between \Delmue\ and \DelSig\ as a function of bar presence.  Figure~\ref{fig:figure9} shows generally no large difference between the two distributions aside from the well-known tendency of bars to inhabit galaxies with larger bulges and thus higher central densities \citep[e.g.,][]{Masters2011}. Any offset for barred \vs unbarred galaxies should be larger for P-bulges, which have intrinsically lower surface brightness and would therefore be inflated more by the presence of bar light.  However, the median value of \Sig\ in barred P-bulges (green points) is only 0.06 dex (16\%) higher than in non-barred P-bulges, which is small.  We also checked for systematic differences between barred and nonbarred galaxies in Figures \ref{fig:figure10}--\ref{fig:figure14} below but see none.  We conclude that the effect of bars on our classification parameters is small.

In summary, Figures~\ref{fig:figure7}--\ref{fig:figure9} collectively imply that \DelSig\ and \Delmue\ both measure central density and that \DelSig\ is therefore a reasonable substitute for \Delmue\ as a bulge classification parameter. This finding agrees with previous works that have also found that central density is closely correlated with bulge type \citep[e.g.,][]{Carollo1999, Fisher2016}.   Unlike \Delmue, \DelSig\ does not require bulge-disk decomposition and can be measured directly from SDSS images out to $z = 0.07$.  Use of  \DelSig\ therefore opens the way to studying bulge properties using the full statistical weight of SDSS.

\section{\Delmue\ and \DelSig\ compared to other structural parameters}
\label{sec:5}

\begin{figure*}
    \includegraphics[width=1.75\columnwidth]{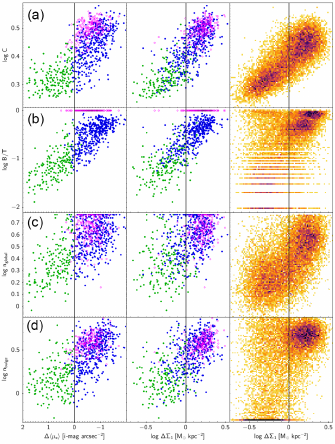}
    \caption{This figure compares four structural parameters \vs\ the structural P-bulge/C-bulge indicators \Delmue\ and \DelSig. The two plots on the left in each row compare \Delmue\ and \DelSig\ for the G09 sample, which includes both central and satellite galaxies.   Green and blue points and magenta circles represent P-bulges, C-bulges and E's as classified by \citetalias{Gadotti2009}. The vertical black lines are the adopted boundaries in \DelSig\ (here) and \Delmue\ (in \citetalias{Gadotti2009}) that divide P-bulges from C-bulges according to each criterion (P-bulges are to the left of the line in each plot). The rightmost plots are the same structural parameters \vs\ \DelSig\ for the mass-limited SDSS \emph{central} sample. SDSS points are weighted by the completeness correction computed in Section \ref{sec:2}; E's are included in the SDSS sample but are not separately marked. The stellar mass, $b/a$, and redshift limits of the G09 sample are log\ \mass/\msun\ > 10, $b/a$ > 0.9, and 0.02 < $z$ < 0.07, while the stellar mass, $b/a$, and redshift limits of the mass-limited SDSS sample are 10.0 < log\ \mass/\msun\ < 10.4, $b/a$ > 0.5 and 0.02 < $z$ < 0.07. The lower massses in SDSS explains why some parts of the diagrams are not populated (e.g., high values of C are missing).  $B/T$ and $n_{\rm bulge}$ for the bulges of G09 galaxies are from the decomposition data in \citetalias{Gadotti2009}, while $B/T$ and $n_{\rm bulge}$ for the galaxies in the SDSS sample are from \citet{Simard2011}. A group of galaxies was originally seen in the upper left corner of the SDSS sample in panel b.   Many are objects for which the fitting procedure misidentified bulges as disks and vice versa \citep[e.g.,][]{Simard2011}. They are omitted from panels b and d. The outlier population in this region of panel b is thereby reduced but still not entirely removed.}
    \label{fig:figure10}
\end{figure*}

\begin{figure*}
    \includegraphics[width=1.75\columnwidth]{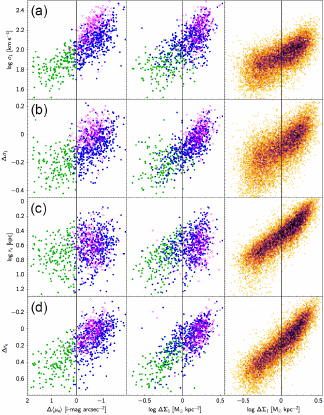}
    \caption{Two additional structural parameters are compared to \Delmue\ and \DelSig.  The format is the same as Figure~\ref{fig:figure10}. The Y-axis is reversed for $r_e$ and $\Delta r_e$ to match the sense of other figures.  The quantity $\sigma_1$ is the velocity dispersion scaled to a circular aperture of radius 1 kpc, which is computed from the SDSS fiber velocity dispersion using the relation $\sigma \propto r^{-0.066}$ according to \citet{Cappellari2006}.   Note the lack of high-$\sigma_1$ galaxies and the lower scatter in the mass-limited (and central) SDSS sample due to the lack of massive galaxies.  Velocity dispersion and effective radius of the galaxies (global) are shown raw and with mass trends removed by fitting and subtracting the ridgeline relations for quenched galaxies \vs\ mass.} The use of such residuals is appropriate when using \DelSig, which also has its mass trend removed.  The scatter is accordingly reduced for \DelSig\ in panels b and d for the G09 sample, which has a large mass range.
    \label{fig:figure11}
\end{figure*}

This section compares \Delmue\ and \DelSig\ to other structural parameters. Many of these have been mentioned in the literature as useful discriminators between P-bulges and C-bulges \citepalias[e.g.,][]{Kormendy2016}.  The comparisons are illustrated in Figures~\ref{fig:figure10}--\ref{fig:figure11}.  Each parameter is shown with three plots.  The left and middle plots show the G09 galaxies colored by bulge types from \citetalias{Gadotti2009}.  The leftmost panel uses \Delmue\ from \citetalias{Gadotti2009}, while the middle panel uses \DelSig.  The far right plot repeats \DelSig\ for the mass-limited SDSS sample from Table~\ref{table:sample} with 10.0 < log\ \mass/\msun\ < 10.4, where SDSS is nearly complete. Ellipticals are included in all plots but are not separately marked in the SDSS sample; in this mass range, they are a minority of galaxies, even in the evolved ridgeline population.

To homogenize the appearance of the plots, the Y-axes are inverted as needed to place quenched galaxies at the top and star-forming galaxies at the bottom. The vertical black lines divide bulge types: galaxies to the left of these lines are classed as P-bulges and galaxies to the right are C-bulges, according to which X-axis parameter is used.  The basic source of all data unless otherwise stated is SDSS DR7 \citep{Abazajian2009}.  Any additional special treatments are described in the captions and the text below.

Before describing the figures in detail, we remind readers of our picture that galaxies are \textit{on average} evolving from low central density to high central density, and thus that there is a \textit{net flow} of galaxies over time in each figure from left to right.  The basis of this point is Figure~\ref{fig:figure1}a, where the narrowness of the \Sig\ locus establishes its nature as an approximate evolutionary track along which \Sig\ and stellar mass both grow with time.  A further implication is that galaxies must also on average be moving across the SV, which is implied by the steadily increasing number of galaxies on the \Sig\ ridgeline from $z = 2.5$ to now \citep{Barro2017}.\footnote{The \Sig\ ridgeline in \citet{Barro2017} includes both star-forming galaxies as well as quenched galaxies.  They are analogous to the star-forming classical bulges (C-SFBs) that are identified later in this paper, i.e. the ``elbow" galaxies shown in Figure~\ref{fig:figure16}.}  The important conclusion is that, if bulge type is based on central stellar density (i.e., \DelSig\ or \Delmue), \emph{at least some pseudo-bulges are evolving to become classical bulges.} This answers a major question set up in the Introduction, namely, how bulge types relate to galaxy evolution.  However, a small caveat is necessary.  Many of the following figures show prominent ridgelines, and it is tempting to interpret these as actual evolutionary tracks, analogous to the giant-branch ridgelines in HR diagrams. Although this is very broadly true, future papers will show that the ridgelines in Figures~\ref{fig:figure10}--\ref{fig:figure14} are in fact composites of \textit{different galaxy sub-populations}, and individual galaxies do not necessarily evolve all the way from the far left to the far right in each diagram.

We start with Figure~\ref{fig:figure10}, which displays correlations with four structural parameters measured from the galaxy images.

\begin{itemize}

\item10a, log $C$ ($R90/R50$):  The agreement between log $C$ (concentration) and \Delmue\ and \DelSig\ is one of the best among all panels, showing few discrepant galaxies off the main relations in the lower-right or upper-left corners.   \DelSig\ (correlation coefficient $r = 0.79$) is a bit tighter than \Delmue\ ($r = -0.75$), but both parameters are well correlated.  Use of log on the Y-axis (and in other panels) matches the use of logs on the X-axis.  Consistent use of logs on both axes makes all structural diagrams look roughly like power laws. The SDSS sample at right shows an extra blob (more accurately, a widening) at log $C \sim 0.44$, \DelSig\ $\sim$ +0.3 that is not seen in the G09 sample (perhaps this is due to the fact that there are too few galaxies in G09 sample). This is an example of a sub-population, and future papers will show that this feature is associated with the smallest quenched galaxies in this SDSS mass range.   Values of $C$ in the SDSS sample do not reach the highest values present in the G09 sample owing to the mass limit at log\ \mass/\msun < 10.4 in SDSS. 

\item10b, log $B/T$: The two left plots use log $B/T$ values from \citetalias{Gadotti2009}. Both are tight, with \DelSig\ ($r = 0.76$) comparable to \Delmue\ ($r = -0.75$). The absence of galaxies above log $B/T$=-0.01 is due to the fact that \citetalias{Gadotti2009} places all high-$B/T$ objects in the elliptical category, which are shown as the magenta circles at the top of the panels.  The SDSS panel uses $B/T$ from \citet{Simard2011}. The discontinuity for the low-$B/T$ galaxies is due to the fact that \citet{Simard2011} does not use 3 decimal digits for $B/T$. A group of galaxies was originally seen in the upper left corner of the SDSS sample.   Many of these proved to be objects in which the fitting procedure misidentified bulges as disks and vice versa \citep[e.g.,][]{Simard2011}. Their $B/T$ were much too high, and they are omitted from this panel (and also from panel d). The outlier population is consequently reduced but is still not entirely eliminated.  Agreement between $B/T$ and \DelSig\ is much poorer overall for SDSS than for the G09 sample, possibly indicating less accurate bulge-disk decompositions in \citet{Simard2011}.

\item10c, log $n_{\rm global}$: All $n_{\rm global}$ indices are single-\Sersic\ fits from the NYU-VAGC. In the G09 plots at left,  \Delmue\ ($r = -0.65$) now is tighter than \DelSig\ ($r = 0.52$), but the scatter in both plots is large.  The SDSS and G09 samples look basically the same, but high-$n$ values are missing from SDSS, a consequence of the $10^{10.4}$ \msun\ upper mass limit. 

\item10d, log $n_{\rm bulge}$: The two left plots use $n_{\rm bulge}$ from \citetalias{Gadotti2009}. Both relations have intermediate tightness, and \Delmue\ ($r = -0.67$) is slightly tighter than \DelSig\ ($r = 0.61$). The SDSS panel uses $n_{\rm bulge}$ from \citet{Simard2011}.  The vertical distribution is very different from \citetalias{Gadotti2009}, there being many more low-$n_{\rm bulge}$ objects in SDSS.  This again is due to the low-mass cut in SDSS.  The tail of outliers in the upper-left corner that is visible in panel b is present here, too.  Even though $n_{\rm bulge}$ < 2 (log $n_{\rm bulge}$ < 0.3) has been used as a prime pseudo-bulge discriminant in the past \citep{Fisher2008}, \citetalias{Gadotti2009} gave it low weight based on poor agreement with his \Delmue. This scatter is visible in the far-left panel using \Delmue\ and is replicated using \DelSig\ in the middle panel. Much of this is probably intrinsic, but an additional factor is that $n_{\rm bulge}$ (and $B/T$) depend on bulge-disk decompositions.  The large scatter for SDSS here parallels the similar scatter in panel b and suggests less accurate bulge-disk decompositions in \citet{Simard2011}.

\end{itemize}

Several conclusions follow from Figure~\ref{fig:figure10}.  The most important is that both \Delmue\ and \DelSig\ show reasonable correlations with other well measured structural parameters and that \DelSig\ correlates comparably to \Delmue\ with these other measures. The second point is that bulge types using \Delmue\ \vs\ \DelSig\ agree well.  It is true that a few galaxies are on different sides of the vertical lines in the two left panels, but this is a detail -- the major features of both diagrams are the same. We therefore feel comfortable in generally denoting P-bulges as low-density bulges and C-bulges as high-density bulges without having to specify which measure, \Delmue\ or \DelSig, we are using.  If it matters, we will be more specific.  Finally, all panels show considerable spread, and because of this there are inevitably going to be discrepant cases choosing bulge types based on different parameters.  This problem is only moderately severe in certain panels in Figure~\ref{fig:figure10} (e.g., panel a, which uses concentration) but it will become very severe in future figures that use spectral indices.  The point is, scatter in these panels means that the two potential bulge-class parameters on the X- and Y-axes disagree, and the larger the scatter, the worse the disagreement.

We turn now to Figure~\ref{fig:figure11}, which adds two more structural parameters, central velocity dispersion $\sigma_1$ and effective radius of the galaxies $r_e$. Each is shown raw and as a residual with mass trend removed. The latter is consistent with the removal of the mass trend in  \DelSig, and using these residual definitions considerably tightens correlations for it.  

\begin{itemize}
  
\item11a, log $\sigma_1$:  \Delmue\ ($r = -0.77$) is tighter than \DelSig\ ($r = 0.67$), but this is due to the fact that the mass trend in $\sigma_1$ has not yet been removed (see panel b). The lack of high dispersions in SDSS is due to the lack of galaxies above $10^{10.4}$ \msun, whereas these are present in the G09 sample. The scattering of low points in all panels reflects the larger fractional errors in SDSS velocity dispersions below 70 km s$^{-1}$.

\item11b, log $\Delta\sigma_1$: The mass trend in $\sigma_1$ is removed by substituting $\Delta\sigma_1$, which is defined as log $\Delta\sigma_1$ = log $\sigma_1$ - 0.338*\mass\ + 1.430. Residual quantities are now used consistently on both axes for \DelSig. \DelSig\ ($r = 0.69$) is now tighter than \Delmue\ ($r = -0.61$), and the range of the SDSS sample on the Y-axis is more consistent with the G09 sample. \Sig\ was shown to correlate closely with central velocity dispersion $\sigma_1$ \citep{Fang2013} through the relations $M$ $\sim$ $\sigma_1^{2}$  and $M$ $\sim$ $\frac{\sigma_1^{2}}{r}$, with all quantities measured within 1 kpc. The good correlation here between $\Delta\sigma_1$ and \DelSig\ is therefore expected. 

\item11c, log $r_e$:  The Y-axis has been reversed in order to match the sense of other diagrams.  The large scatter in \Delmue\ and \DelSig\ reflects systematic residuals with mass.  This effect is less evident in the SDSS panel since the mass range is smaller. 

\item11d, log $\Delta r_e$: The Y-axis is again reversed. The mass trend in $r_e$ is removed by substituting $\Delta r_e$, which is defined as log $\Delta r_e$ = log $r_e$ - 0.535*\mass\ + 5.175. Residual quantities are again now used consistently on both axes.  \DelSig\ for the G09 sample tightens dramatically, as expected since \Sig\ and $r_e$ are closely correlated at fixed mass for star-forming galaxies with \Sersic\ in the range $n = 1$-2  \citep[e.g.,][]{Barro2017}; elliptical galaxies, which have higher \Sersic\ indices and therefore a different relationship between \Sig\ and $r_e$, lie systematically low.  

\end{itemize}

Figure~\ref{fig:figure11} extends the conclusion from Figure~\ref{fig:figure10} that \Delmue\ and \DelSig\ compare well with each other and with other well-measured structural parameters.  For maximum tightness, use of \DelSig\ requires that mass trends be removed from both coordinates.

\section{\Delmue\ and \DelSig\ compared to stellar-population parameters}
\label{sec:6}

\begin{figure*}
    \includegraphics[width=1.75\columnwidth]{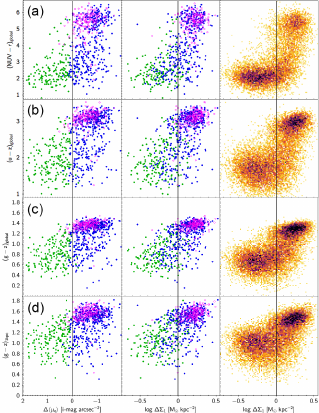}
    \caption{Four colors that are sensitive to star-formation rate, stellar age, and dust content are plotted \vs\ \Delmue\ and \DelSig.  Three are global, the last is central.  The format is the same as in Figures~\ref{fig:figure10} and \ref{fig:figure11}.  All colors have been corrected for dust using the global estimates of \citet{Oh2011}.  The SDSS sample is smaller than the main sample by $\sim$1,000 galaxies due to the extra requirement that spectroscopic S/N > 10.  Inner 1-kpc color $(g-z)_{\rm 1kpc}$ is computed from the surface brightness profile obtained from the SDSS database (the color $g-z$ has been used since $u$ is too noisy through a 1 kpc aperture).  Note the elbow-shaped distributions in $NUV-r$ and (to a lesser extent) in $u-z$. It is shown in the next panels that galaxies in the elbow have high central density yet also high star formation rate.  These objects figure prominently in the discussion, and we have named them star-forming classical bulges (C-SFBs; see also Figure~\ref{fig:figure16}).}
    \label{fig:figure12}
\end{figure*}

\begin{figure*}
    \includegraphics[width=1.75\columnwidth]{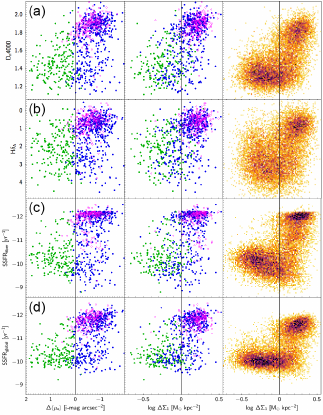}
    \caption{Three stellar population indices from SDSS spectra plus a fourth index showing global specific star-formation rate from \citet{Brinchmann2004} are compared to \Delmue\ and \DelSig.  The format is the same as in Figures~\ref{fig:figure10} and \ref{fig:figure11}.  The presence of the elbow in the first three panels confirms the presence of ongoing star formation and young stellar ages \textit{in the centers} of star-forming classical bulges (C-SFBs).  The fourth panel agrees with $NUV-r$ in Figure \ref{fig:figure12} and establishes that young stars are present \textit{throughout}, not just in the inner parts.  The \textit{downward trend} from left to right for star-forming galaxies in panel c indicates even higher star-formation rates at higher central densities.}  
    \label{fig:figure13}
\end{figure*}

\begin{figure*}
    \includegraphics[width=1.75\columnwidth]{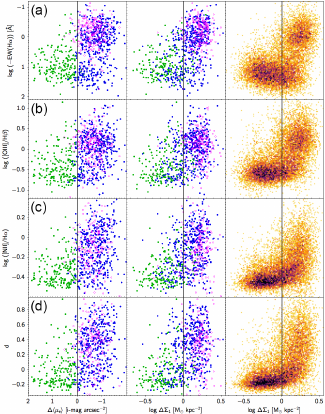}
    \caption{Four emission-line ratios compared to \Delmue\ and \DelSig.  The format is the same as in Figures~\ref{fig:figure10} and \ref{fig:figure11}.  The H$\alpha$ equivalent width is multiplied by -1 and the y-axis is reversed to maintain the same sense as in other figures. (Taking logs loses roughly 2\% of galaxies, which have intrinsically positive EWs near zero.)  The quantity $d$ is the distance of a galaxy from the line dividing star-forming galaxies from AGNs according to \citet{Kauffmann2003c}; its equation is given in the text.  These emission-line data confirm the presence of ongoing star formation at the centers of star-forming classical bulges (C-SFB), in agreement with the absorption indices and  $\rm SSFR_{\rm fiber}$ in  Figure~\ref{fig:figure13}.  The downward tilt in EW(H$\alpha$) signals higher \textit{ongoing} star formation in elbow galaxies.}
    \label{fig:figure14}
\end{figure*}

The previous section showed correlations between \Delmue\ and \DelSig\ and other structural parameters.  These correlations are quite linear in log-log coordinates, and the distribution of points along most relations is fairly uniform, there being no separate clump due to quenched galaxies.  Figures~\ref{fig:figure12}--\ref{fig:figure14} now compare these two variables to stellar population parameters. The format of the figures and samples used are the same as in Figures~\ref{fig:figure10} and \ref{fig:figure11}.  Parameters are grouped into categories by type.  Photometric indices characterizing stellar age and star-formation rate are shown in Figure~\ref{fig:figure12}, followed by stellar-population spectral indices derived from SDSS spectra\footnote{As mentioned in Section~\ref{sec:2} describing the sample, a signal-to-noise cut of S/N > 10 per pixel was applied in choosing the spectroscopic sub-sample, which caused the loss of about 1,000 additional galaxies out of 12,000 (see Table 1).   We have verified that these galaxies are mostly lost because they are dim, but they are otherwise relatively uniformly distributed as a function of \DelSig\ and color in Figures~\ref{fig:figure12}--\ref{fig:figure14}.  Their loss does not therefore substantially distort the sample. No additional cuts are applied for emission-line S/N.} of the central regions in Figures~\ref{fig:figure13} and \ref{fig:figure14}. Three AGN-related indices complete the list. Some Y-axes are inverted to put quenched galaxies at the top.

Figure~\ref{fig:figure12} shows four color indices that are diagnostic of stellar age, star-formation rate, and/or dust. The first three are global, the last one is for the inner 1 kpc only. 

\begin{itemize}
  
 \item12a, $NUV-r$: This UV index is sensitive to ongoing star formation with an averaging time of a few tens of millions of years \citep[e.g.,][]{Yesuf2014}.  The quenched clump at upper right is under-populated because red galaxies tend to be missed in GALEX on account of their low UV flux.  The main new feature in this plot is the pronounced ``elbow'', which is visible in both the G09 and SDSS samples.  The elbow was discovered previously in SDSS galaxies by \citet{Fang2013} using $NUV-r$ \vs\ \DelSig\ and shown to exist out to $z=3$  by \citet{Barro2017} using specific star-formation rate instead of $NUV-r$ \citep[see also][]{Mosleh2017,Lee2018}. 
 
 \end{itemize}
 
 We call attention to the dramatically different appearance of this panel from previous panels in Figures~\ref{fig:figure10}--\ref{fig:figure11}, which used structural variables and were generally monotonic. The difference signals a divergence between structural properties and stellar population properties, which will be repeated in all subsequent spectral parameters.  Objects in the clump at upper right are quenched galaxies with high central density.  Objects at lower left are star-forming with low central density. The objects in the elbow are aberrant in having high star formation rates despite having high central density.  We have named these objects star-forming classical bulges (C-SFBs), and their location is also highlighted in Figure~\ref{fig:figure16}.  These objects seem poised to enter the green valley, and their properties may therefore provide a clue to the mechanics of quenching. The ridgeline of star-forming galaxies at the bottom of the distribution is also strikingly flat, suggesting no large trend in \textit{global} specific star formation rate with central density as long as galaxies are star-forming.
 
 The preceding paragraph has defined a new subclass of C-bulges, but it is seen from the G09 plots in Figure~\ref{fig:figure12} that the exact membership in this class is not quite the same depending on whether C-bulges are defined using \Delmue\ or \DelSig.  While \DelSig\ is preferred because it is used for the larger SDSS sample, we stress that the new sub-class C-SFBs is basically just a \textit{useful name} that calls attention to the fact that the relation between the structural variables and the stellar population variables is highly non-linear, which is true regardless of whether \DelSig\ or \Delmue\ is used.  The exact membership in this class is not important in what follows. 

\begin{itemize}

 \item12b, $(u-z)_{\rm global}$. The plot of $(u-z)_{\rm global}$ resembles that of $NUV-r$ but with lower dynamic range.  The elbow is still present but is slightly less prominent. The quenched sequence at the top of the SDSS distribution is tilted slightly upward, suggesting an age gradient within the quenched population. Unlike $NUV-r$, the ridgeline of star-forming galaxies tilts slightly downwards toward higher \DelSig. Overall, these diagrams resemble plots of \Sersic\ index and/or concentration \vs\ global $(u-r)$ in \citet{Driver2006,Baldry2006,Ball2008}, which with the benefit of hindsight look rather elbow-shaped.

\end{itemize}

The next two panels compare central \vs\ global colors in the same color index. Since $u$ is too noisy through a 1-kpc aperture, $g-z$ is used.

\begin{itemize}

\item12c, $(g-z)_{\rm global}$: The dynamic range of this color index is less than $NUV-r$ or $u-z$.  The upward tilt of quenched galaxies that was seen in $u-z$ remains visible, but the elbow looks weaker because the color of the C-SFB population is quite red in this index.  It appears that some C-SFB objects have moved up into the green valley, or even merged with the quenched clump using this color.  The combination of red  $g-z$ plus blue $NUV-r$ could mean dust \citep[e.g.,][]{Williams2009, Patel2011, Brammer2011} or composite stellar populations \citep{Wang2017}.  \DelSig\  appears slightly tighter than \Delmue\ in the G09 sample.

\item12d, $(g-z)_{\rm 1kpc}$:  This $g-z$ is measured within 1 kpc whereas the dust correction from \citet{Oh2011} is global, and so there is the possibility of mismatch.  The total dynamic range is smaller than $(g-z)_{\rm global}$, which says that color gradients are larger in star-forming galaxies than in quenched galaxies (centers are slightly redder). Whether this is due to older stars or more dust cannot be answered without more data.

\end{itemize}

The main result from Figure~\ref{fig:figure12} is to highlight evidence for a high-density star-forming population near the elbow of $NUV-r$ and other colors, which we have termed the C-SFB population. Such a population was seen before in \citet{Fang2013,Barro2017,Woo2017,Mosleh2017,Lee2018} using \textit{global} indices.  An attempt using $g-z$ in Figure~\ref{fig:figure12}d to see whether young stars also extend to the \textit{centers} of these galaxies was inconclusive.  Figures~\ref{fig:figure13} and \ref{fig:figure14} will test this idea using central SDSS indices.

\begin{itemize}

\item13a, D$_n$4000:  D$_n$4000 is a commonly used age indicator that averages star formation over timescales of a few hundred million years \citep[e.g.,][]{Yesuf2014}.  In Figure~\ref{fig:figure13}a, \Delmue\ and \DelSig\ scatter nearly equally in the G09 sample, and the C-SFB elbow is very strong in both, confirming that young central stars are present.  Additional panels with other indices below confirm this.  It is interesting to compare the large SDSS population in the right panel with the corresponding panel for $NUV-r$ in Figure~\ref{fig:figure12}a.  The elbow is prominent in both, but D$_n$4000  actually \textit{declines} with \DelSig\ in star-forming galaxies, indicating \textit{even younger stars at the centers} of C-SFB galaxies (the SDSS fiber spans approximate 1-4 kpc at these distances). This appears in subsequent figures and agrees with similar findings by \citet{Woo2019}, who found younger stars in the centers of denser star-forming galaxies (see below). \citet{Kauffmann2003b} plotted D$_n$4000 \vs\ \textit{effective} stellar mass density, $\mu_*$, rather than \DelSig\ as here.  The same two clumps of star-forming and quenched galaxies were seen coexisting at high density, but the decline in D$_n$4000 among the star-forming population was not apparent.  This may be because they used absolute density, not a residual, for a sample that contained a wide range of stellar mass.

\item13b, H$\delta_{\rm A}$: This index is corrected for emission and is a sensitive young-star indicator with a response time comparable to $NUV-r$ \citep{Yesuf2014}. \Delmue\ and \DelSig\ both scatter broadly relative to H$\delta_{\rm A}$ in the G09 sample. In SDSS, the quenched peak is strong and tight, but star-forming galaxies scatter widely. The latter effect hints at different central star formation histories even while galaxies are on the main sequence.  Known processes that can cause H$\delta_{\rm A}$ to differ from longer-timescale indices (like D$_n$4000) include starbursts and rapid quenching \citep[e.g.,][]{Dressler1983,Zabludoff1996,Quintero2004,Yesuf2014}. However, independent of scatter, the general star-forming population again tilts downward \vs\ \DelSig, replicating D$_n$4000 and supporting the presence of relatively younger stars at the centers of C-SFBs. The large scatter in H$\delta_{\rm A}$ in star-forming galaxies compared to D$_n$4000 and other central indices (see below) is another new finding and is not understood. 

\item13c, $\rm SSFR_{\rm fiber}$: This is the specific star-formation rate within the fiber from \citet{Brinchmann2004}. A major surprise is the \emph{very} strong downward trend with \DelSig, signaling much larger SSFR in the centers at high density and supporting and amplifying previous results from D$_n$4000 and H$\delta_{\rm A}$ and from $\rm SSFR_{\rm global}$ in the next panel.   

\item13d, $\rm SSFR_{\rm global}$: This is the specific star-formation rate for the entire galaxy from \citet{Brinchmann2004}. It combines the measurement of H$\alpha$ flux in the fiber with additional color information from the outer parts. Its morphology generally matches that of the other global index, $NUV-r$ in Figure~\ref{fig:figure12}a. The C-SFB population in the elbow is prominent, signaling strong ongoing global star formation, and the ridgeline is much more level than in D$_n$4000 or H$\delta_{\rm A}$ and more like $NUV-r$. 

\end{itemize}

In total, Figures~\ref{fig:figure12} and \ref{fig:figure13} strongly establish the presence of young stars and ongoing star formation both globally and at the centers of C-SFB galaxies.  The high star-formation rates in these galaxies are therefore a general phenomenon not confined to the outer parts (i.e., we are not seeing dead, red bulges and blue outer disks). There is evidence from D$_n$4000,  H$\delta_{\rm A}$, and $\rm SSFR_{\rm fiber}$ of even stronger star formation activity in the centers of C-SFB elbow galaxies than in the centers of low-\Sig\ galaxies farther from the quenched ridgeline. This finding is consistent with the results of \citet{Woo2019}, who study age and star-formation-rate gradients in star-forming SDSS galaxies as a function of \DelSig. Galaxies with high \DelSig\ tend to have strong age gradients with younger stellar populations and higher specific star-formation rates in their centers, while lower-\DelSig\ galaxies have older centers with lower specific star-formation rates.  \citet{Woo2019} study gradients whereas we study absolute central values, but the results are consistent. 

Finally, Figure~\ref{fig:figure14} adds four more emission-line indices that shed further light on star formation and AGNs. 

\begin{itemize}

\item14a, EW(H$\alpha$): This index has not been corrected for dust, but we have verified that the basic morphology remains unchanged if the reddening corrections of \citep{Oh2011} are used (note that EWs here are multiplied by -1). The morphology is similar to D$_n$4000 in Figure~\ref{fig:figure13}a. \Delmue\ and \DelSig\ in the G09 sample look similar, and both exhibit a strong elbow, as does SDSS.  The quenched clump in SDSS is broadened due to the presence of Seyfert and LINER emission, which varies from galaxy to galaxy.  The downward tilt along the horizontal branch duplicates similar trends in D$_n$4000, H$\delta_{\rm A}$, and  $\rm SSFR_{\rm fiber}$, but the very short timescale of H$\alpha$ directly implies higher \textit{ongoing} star formation in elbow galaxies, not just younger average age. 

\item14b [OIII]/H$\beta$: This is one of two line ratios used to divide star-forming galaxies and AGNs in the BPT diagram. \Delmue\ and \DelSig\ in the G09 sample look similar, and both exhibit a strong elbow in the C-SFB population, which is also seen in the SDSS sample. The ratio is flat at the star-forming value into the elbow, indicating ongoing star-formation there, shifting to the AGN value for galaxies in the quenched clump.  

\item14c [NII]/H$\alpha$: This is the other line ratio used to divide star-forming galaxies from AGNs in the BPT diagram. The plots resembles [OIII]/H$\beta$ except that [NII]/H$\alpha$ scatters more widely in quenched galaxies, and [NII] appears to increase faster than [OIII] as galaxies evolve into the green valley.  

\item14d $d$: The quantity $d$ is the slanting distance of a galaxy from the dividing line between star-forming and AGN galaxies in the BPT diagram defined by \citet{Kauffmann2003c} and is computed as $d$ = log ([OIII]/H$\beta$) $-$ 0.61/(log ([NII]/H$\alpha$) - 0.05) - 1.3. The diagram looks like an average of [OIII]/H$\beta$ and [NII]/H$\alpha$, as expected.  

\end{itemize}

We now summarize the major conclusions from  Figures~\ref{fig:figure10}--\ref{fig:figure14}.  We stress again that the SDSS sample is for \textit{central} galaxies only with log\ \mass/\msun\ > 10.0 and the behavior of satellites may be different \citep{Woo2017}.

First, \Delmue\ and \DelSig\ appear to characterize the structural state of bulges similarly in that both measure bulge prominence via central density.  Although the left-hand and middle panels of these figures differ in detail, they are broadly similar.  Since \Delmue\ is one of the standard parameters used to classify P-bulges and C-bulges \citepalias[e.g.,][]{Gadotti2009, Fisher2016}, it follows that \DelSig\ is also a serviceable indicator of bulge structure, with the additional advantages that it does not require bulge-disk decomposition and it can be measured out to $z = 0.07$ in SDSS images \citep{Fang2013} and out to $z = 3$ in \textit{HST} images \citep{Barro2017}.  We have therefore met one of the major goals of this paper, to compare \DelSig\ to at least one other classical bulge structure indicator, and we have shown that it compares favorably and measures similar bulge properties. We also note that \DelSig\ has had the mass trend removed, and therefore for consistency other mass-dependent parameters should also have their mass trends removed before comparing to it. 

Second, the different panels of Figures~\ref{fig:figure10}--\ref{fig:figure14} exhibit a wide variety of different morphologies that suggest a wealth of information yet to be unlocked with additional analysis.  This is reinforced by the discovery, to be discussed in future papers, that the broad, fuzzy loci in Figures~\ref{fig:figure10}--\ref{fig:figure14} are actually comprised of sub-populations in different evolutionary stages. This is the reason we have warned against over-interpreting the entire ridgelines as evolutionary tracks.  Instead, it is the various sub-populations \textit{within} these patterns that are the tracks, as will be shown in future papers.

Third, the locus of star-forming galaxies is rather flat \vs\ \DelSig\ using global parameters like $NUV-r$, $u-z$, and $\rm SSFR_{\rm global}$ but trends downward at high \Sig\ using central parameters like D$_n$4000, H$\delta$, EW(H$\alpha$), and $\rm SSFR_{\rm fiber}$.\footnote{We have verified that this effect is not due to a fiber aperture effect by seeing no redshift dependence in any plot.} This shows conclusively that not only is global star-formation high in C-SFB elbow galaxies but central star formation is even higher. This finding of younger stars and higher star formation at the centers of high-density C-SFBs agrees with measurements of stellar-population gradients by \citet{Woo2019}.

The final point, mentioned above, is that correlations involving purely structural parameters (including \Delmue\ and \DelSig) seem to be rather straight in log-log space whereas correlations that mix stellar-population variables with \Delmue\ and \DelSig\ are elbow-shaped.  We have verified separately that the elbow objects are substantially the same in all diagrams. Together, these results suggest that central galaxies in this mass range possess a well-correlated set of structural parameters and a separate set of well-correlated stellar-population parameters and that elbow relations result when mixing parameters from the two classes.
 
\section{Discussion}
\label{sec:7}

\subsection{Comparison to previously measured bulge types}
\label{sec:7.1}

Our most important finding is that star-formation rates do not correlate perfectly with central structure -- galaxies with \DelSig\ < 0 are all star-forming, whereas galaxies with \DelSig\ > 0 are a mixture of quenched and star-forming (the ``elbow'' pattern).  This pattern persists even when even more global structural parameters, such as concentration, $B/T$, $n_{\rm global}$, and $r_e$, are used.  Finally, different indices are quite consistent: all structural parameters show roughly linear relations against \DelSig, whereas all star-forming indices show elbows.  

As shown in the left and middle panels of Figures~\ref{fig:figure10}--\ref{fig:figure14}, these results agree very well with \citetalias{Gadotti2009}, as might be expected since our \DelSig\ bulge calibration is closely modeled on his parameter \Delmue.  However, the bulge-type literature in general has used a much wider range of bulge-classification parameters, and to compare to them, we use the summary of correlations by \citetalias{Fisher2016}.  Several conclusions are at variance with the very clear trends in Figures~\ref{fig:figure10}--\ref{fig:figure14}.  Broadly speaking, sources agree that the majority of P-bulges are low-density and high star-forming and that the majority of C-bulges are high-density and low star-forming, but difficulties arise when trying to make sense of the outliers.

To illustrate, we select four findings from \citetalias{Fisher2016} and add some comments.  We primarily rely on the SDSS sample but refer occasionally to the G09 sample when needed.

\begin{itemize}

\item ``Though classical bulges are rarely found to be blue, pseudo-bulges are often red.'' Neither of these conclusions agrees with our data.  If classical bulges are defined as galaxies that lie to the right of the vertical lines in Figures~\ref{fig:figure10}--\ref{fig:figure14}, it is seen that a substantial portion of them are blue and star-forming. More quantitatively, using the mass-limited SDSS sample and setting aside the 642 elliptical galaxies leaves 5588 C-bulges.  Dividing them at D$_n$4000 = 1.6 gives 3238 red galaxies and 2350 blue galaxies.  Thus, 42\% of \textit{central C-bulges in the mass range 10.0 < log\ \mass/\msun\ < 10.4} are blue.  It is likely that this fraction varies with mass and environment: more massive galaxies are redder and more quenched than the SDSS sample \citep[e.g.,][]{Baldry2006,Ball2008}, and including satellites would also add more red galaxies. Nevertheless, the fraction of 42\% in our sample does not really merit the word ``rare".  

The second part of the sentence also does not agree, as pseudo-bulges in the SDSS sample are virtually never red.  Again, the statement may be sample-dependent, as low-\DelSig\ galaxies are redder when they are satellites \citep{Woo2017} and also when they are massive \citep{Ball2008}.  Both satellites and massive galaxies have been pruned from the SDSS sample but not from the G09 sample, and more red P-bulges may indeed be present there.  The tentative conclusion is that the detailed distributions of galaxies in these diagrams may depend on both mass and environment \citep[cf.][]{Ball2008} and that general conclusions should be carefully qualified. 

\item ``If a bulge is star forming (and there is no interaction present) this is very good evidence that the bulge is a pseudo-bulge, but when the bulge is not star forming this does not imply the bulge is classical.''  Again both of these conclusions disagree with our data.  In the SDSS sample, 29\% of star-forming bulges are C-bulges in this mass range, which means that the P-bulge prediction would be wrong nearly one-third of the time, not really a ``very good" prediction.  As for non-star-forming bulges, virtually \textit{all} are classical in the SDSS sample, contrary to \citetalias{Fisher2016}.  However, the number of P-bulges among red galaxies would be increased by including satellites, so the environmental dependence of the distributions may again be an issue.

\item ``Though pseudo-bulges are rarely found to have high $\sigma$, classical bulges may have either high or low $\sigma$.'' This statement generally agrees with our data, especially if one imagines adding more massive galaxies to the SDSS sample in Figure~\ref{fig:figure11}a.  However, we have argued that use of a mass-corrected residual $\Delta\sigma_1$ is more appropriate than the use of an absolute $\sigma_1$.  

\item ``Bulges that consist of both a thin, star-forming pseudo-bulge and a hot-passive classical bulge are very likely present in some galaxies.''  This statement is made in the context of so-called \textit{composite bulges}, which are objects that exhibit properties of both bulge types \citep[e.g.,][]{Erwin2015}. Such objects might be an intriguing way to account for the properties of C-SFBs, i.e., elbow galaxies that have high central density yet high central star formation.  However, we have checked this possibility using the sample of composite bulges in \citet{Erwin2015} and find that few of them actually host active central star-formation (they are mostly S0's). Moreover, the elbow phenomenon strongly exists in $NUV-r$ as well as the SDSS indices (Figure~\ref{fig:figure12}a), and $NUV-r$, being global, would not be much affected by star formation in a pseudo-bulge. It thus appears that the division between active and passive C-bulges is a global phenomenon, not one that is associated with the presence or not of a composite bulge.

\end{itemize}

The previous bullets have highlighted instances of disagreements between this paper and the findings in \citetalias{Fisher2016}. However, the problem of classifying bulges is more general and has been noted in several works \citepalias[e.g.,][]{Gadotti2009,Kormendy2016}.  On reflection, we think that most disagreements in classifying \emph{individual} galaxies can be chalked up to four causes: a certain amount of noisy data, reliance on hard-to-measure and possibly inconclusive quantities like $n_{\rm bulge}$, failure to use mass-corrected residual quantities consistently, and failure to recognize the fundamentally non-linear (elbow-shaped) relation between structure and stellar populations, which give discrepant classifications for elbow galaxies.  Discrepancies in \emph{broader trends} may arise from these effects as well as differences in the mass ranges and environmental densities used.  We hope to explore these second-parameter effects in future papers.

\subsection{Frequency of bulge types \vs\ stellar mass}
\label{sec:7.2}

\begin{figure}
    \includegraphics[width=\columnwidth]{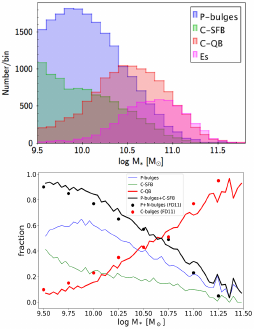}
    \caption{Top panel: The number of SDSS galaxies of different bulge types in 0.2-dex mass bins.  P-bulges are defined as galaxies in the lower-left quadrant of Figure~\ref{fig:figure16}, C-SFBs are the star-forming classical bulges in the lower-right quadrant, and C-QBs are quenched galaxies in the upper-right quadrant (E's included). Magnitude incompleteness corrections are applied to all numbers.  Bottom panel: Fractions of various bulge types \vs\ mass, with ellipticals now excluded.  The black and red dots are analogous data from \citet{Fisher2011}.  Good agreement is achieved if it is assumed that the FD P-bulges comprise mostly star-forming bulges while their C-bulges are mostly quenched, i.e., that their bulge-typing criteria weight stellar-population properties more than central density.}
    \label{fig:figure15}
\end{figure}

Another comparison to previous work is the frequency of bulge types as a function of stellar mass. Figure~\ref{fig:figure15} repeats this plot from \citet{Fisher2011} with our SDSS data added.  The lower limit is set to log\ \mass/\msun\ = 9.5, as SDSS magnitude incompleteness corrections are large below that.  No correction is made for fiber targeting incompleteness (Section~\ref{sec:2}), but if this is a function of apparent magnitude only, then the resulting distributions are at least relatively correct.  For bulge types, we use the three populated quadrants shown in Figure~\ref{fig:figure16}: 1) lower left, P-bulges (\DelSig\ < 0); 2) lower right, star-forming C-bulges (C-SFBs); and 3) upper right, quenched C-bulges (C-QBs).  Ellipticals are included as a fourth category in the upper panel of Figure~\ref{fig:figure15} using P(Ell) > 0.65 from \citet[][]{Huertas-Company2011} (see Section \ref{sec:2}) but are not included in the lower panel.  We cannot distinguish between N-bulges and P-bulges, and so our P-bulge category lumps them together. 

The top panel of Figure~\ref{fig:figure15} shows the number of galaxies per 0.2 dex mass bin. As expected, mass trends are strong, with P-bulges concentrated at lower masses, C-QBs at intermediate masses, and ellipticals at high masses.  Somewhat unexpected is the similarity of the C-SFB bulges to P-bulges.  One might have expected them to be intermediate between P-bulges and C-QBs, but they are very close to P-bulges, suggesting a close evolutionary connection.  

The bottom panel shows fractions of the three bulge types (minus ellipticals) \vs\ mass. Black and red dots show analogous data points from \citet{Fisher2011}.  The black points are for the galaxies they call P+N-bulges, and the red dots are for the galaxies they call C-bulges.  Interestingly, good agreement is achieved if their P-bulges are identified with our P+C-SFBs and their C-bulges are identified with our C-QBs. A hypothesis is that the net criteria used by \citetalias{Fisher2016} weight stellar-population-related properties more heavily than central density.  That would mean assigning types by cutting Figure~\ref{fig:figure16} horizontally through the GV rather than vertically through the SV.  We notice that no matter which criteria we choose to identify P-bulges, there are still many massive P-bulges (more than 20 percent at log\ \mass/\msun\ = 11.0). Given that P-bulges may be converted to C-bulges via mergers, previous authors \citep[e.g.,][]{Kormendy2010b} have highlighted the continued existence of these galaxies. We reinforce this problem here, and \citet{Chen2019} may help explain this from a theoretical point of view.

\subsection{Are bulge properties bimodal?}
\label{sec:7.3}

We turn now to the topic of ``bimodality'', which might potentially play an important role in deducing the origins and evolution of bulges. Much of the literature on bulge types claims that P-bulges and C-bulges are ``bimodal'', which according to the strict definition of the word means a population showing two separate peaks.  A search of the literature reveals only one structural parameter historgram that is convincingly bimodal, a histogram of $n_{\rm bulge}$ that shows two peaks in \citet{Fisher2010}.  But, as the authors say, the sample in that paper is overweighted by Virgo Cluster galaxies, which might tend to produce a false peak at high $n_{\rm bulge}$. In fact, a later version of this histogram  with more galaxies does not display two separate peaks but rather one smooth (if noisy) distribution \citepalias{Fisher2016}.  P-bulges are clustered at one end, and C-bulges are clustered at the other, but there is no strong feature \textit{in the histogram itself} that signals two distinct populations. 

Three points are relevant.  First, it is necessary to employ large and representative samples when testing histograms for bimodality -- overweighting the Virgo Cluster may create a false peak where none exists.  Second, the fact that a second parameter varies systematically from one end of a distribution to the other does not mean that the base population is \textit{bimodal}, it merely says that two variables are correlated.  Third, genuine bimodality would be significant because it might signal that P-bulges and C-bulges are formed by separate mechanisms. In fact, as noted in the Introduction, it is a common view that P-bulges are formed by secular evolution while C-bulges are formed by mergers (or perhaps by violent disk instabilities when galaxies were young and gas-rich) \citepalias[e.g.,][]{Fisher2016, Kormendy2016}.  The existence of two clearly separate peaks in any property might therefore support this picture.

\begin{figure}
    \includegraphics[width=\columnwidth]{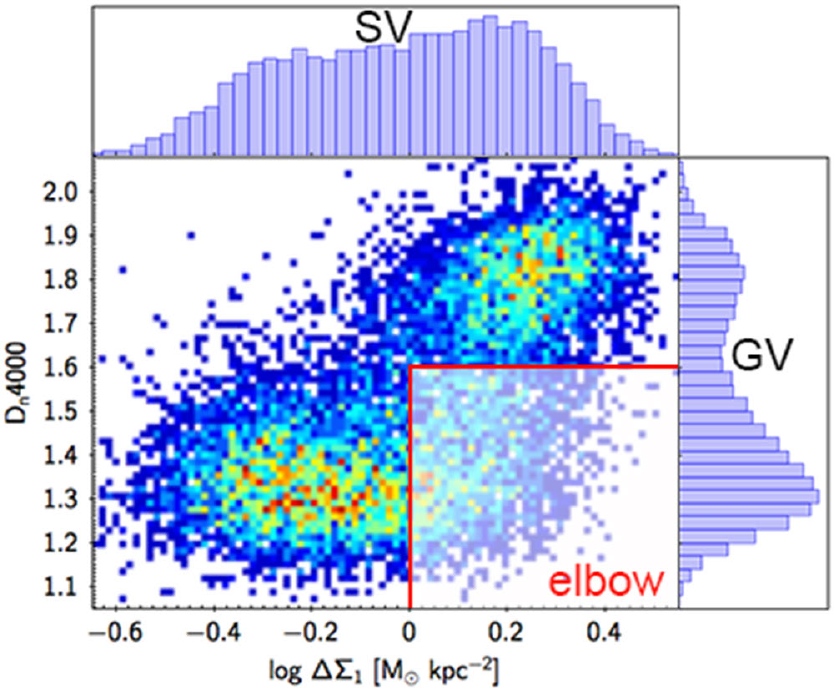}
    \caption{D$_n$4000 \vs\ \DelSig\ and their histograms for our mass-limited SDSS sample of central galaxies in the range 10.0 < log\ \mass/\msun\ < 10.4. Points are color-coded by the number density weighted by the magnitude completeness correction. The dip in the D$_n$4000 histogram is the familiar green valley (GV).  The dip in the \DelSig\ histogram is the structural valley (SV, cf. Figures~\ref{fig:figure1}, \ref{fig:figure5}, and \ref{fig:figure6}).  Even though each axis shows two peaks, the makeup of the peaks is not the same because of the existence of the galaxies at the elbow (shaded region to lower right). Elbow galaxies are spectrally P-bulges but structurally C-bulges, which demonstrates the different nature of the GV and SV -- they are not defined by the same objects.}
    \label{fig:figure16}
\end{figure}

An important insight from Figures~\ref{fig:figure12}--\ref{fig:figure14}  is that bimodality for bulges should properly be considered \textit{simultaneously} in both spectral and structural space. No single histogram, whether it use a structural or a spectral variable, can convey the full 2-D parameter distribution of these objects. This is illustrated in Figure~\ref{fig:figure16}, which replots D$_n$4000 \vs\ \DelSig\ for SDSS central galaxies with masses in the range 10.0 < log\ \mass/\msun\ < 10.4. Histograms of the distributions in \DelSig\ and D$_n$4000 are on the X- and Y-axes. It is true that both histograms are bimodal.  The D$_n$4000 distribution shows the well known dichotomy between star-forming and quenched galaxies, separated by the green valley. The \DelSig\ histogram is also (weakly) bimodal.  This is the same feature that was highlighted in Figures~\ref{fig:figure5}--\ref{fig:figure6}, which we dubbed the \textit{structural valley}.\footnote{The bimodality is weaker here than in Figure~\ref{fig:figure6} because the mass range here is log\ \mass/\msun\ = 10.0 to 10.4 inclusive.}  But these histograms in X and Y are not the most effective way to demonstrate the true bimodality of the population -- that is most clearly shown by the two islands \textit{in two dimensions}, and thus the bimodality in question is neither purely structural nor purely spectral but a combination of the two. The final point is that the objects that comprise the two peaks in D$_n$4000 are not exactly the same as the objects that comprise the two peaks in \DelSig.  That is because the elbow galaxies (shaded region) are members of the P-bulge peak based on star-formation rate but are members of the C-bulge peak based on structure. Hence, a one-dimensional classification system using only structural data, as we have employed here, will necessarily give an incomplete and confusing picture -- the population needs to be modeled in both spectral and structural space simultaneously for a full understanding.

These conclusions were already evident from plots that mixed structural variables and spectral variables in \citet{Kauffmann2003b,Driver2006,Baldry2006,Bell2008} but are now are clearer using the larger array of spectral indices and the mapping onto structural bulge types in Figures~\ref{fig:figure12}--\ref{fig:figure14}.     

\subsection{Bulge types in relation to galaxy quenching}
\label{sec:7.4}

A final point is the close connection between the results here to previous papers on \Sig\ in galaxies.  It is now clear that the ``elbow'' seen in  Figures~\ref{fig:figure12}--\ref{fig:figure14} is the same elbow pattern in SDSS galaxies seen by \citet{Fang2013} in which galaxies with low \DelSig\ have high star formation but galaxies with high \DelSig\ have a wide range of star formation rates.  \citet{Barro2017} and \citet{Lee2018} showed that this elbow pattern is ancient and extends back to at least $z = 3$ in CANDELS. It is thus a deeply ingrained feature of how (central) galaxies fade in our Universe. A major result of the present work is to show how bulge types -- P-bulges, C-bulges, and E's -- map onto this pattern. If the local mapping is universal, distant galaxies would also exhibit P-bulge and C-bulge properties similar to nearby galaxies. If our picture that galaxies are evolving from low \DelSig\ to high \DelSig\ today is correct, it would follow that the entire sequence from N-bulge to P-bulge, C-bulge, and finally to E's is an evolutionary progression that exists at all redshifts.

\citet{Fang2013} attempted to explain the non-linear correlation between star formation rate and \DelSig\ in terms of quenching by  black hole feedback.  Their picture was motivated by the relations $M_{BH} \sim \sigma^4$ and $\Sigma_1 \sim \sigma_1^2$, the latter also measured by \citet{Fang2013}.  Thus $M_{BH} \sim \Sigma_1^2$, and central density \Sig\ becomes an indicator of BH mass. According to this picture, galaxies build bulges and black holes together before they quench. C-SFBs in the elbow are star-forming galaxies at the tipping point, while C-bulges on the vertical branch are former C-SFBs in which BH feedback is finally having a measurable dampening effect on star formation rate. In other words, the elbow-shaped distributions are due to the fact that structure and stellar populations \emph{evolve differently near quenching}.  Remaining questions are why galaxies tend to quench at a particular mass and why the quenched ridgeline in \Sig\ \vs\ \mass\ has the observed slope and zero point that it does (Figure~\ref{fig:figure1}a). These topics are addressed in a companion paper to this one on how black holes might quench galaxies by transferring energy to the hot gas in their dark halos \citep{Chen2019}. 

\section{Conclusions}
\label{sec:8}

In this paper, we study the relationship between the stellar mass surface density within 1 kpc, \Sig, to the nature of galactic bulges in SDSS galaxies with $0.02 < z < 0.07$ and log\ \mass/\msun\ > 10.0. The goals are to establish how \Sig\ relates to other bulge classification parameters and to see if it can be used to identify C-bulges and P-bulges in SDSS.  A residual parameter \DelSig\ is defined by removing the mass trend from \Sig.  Since \DelSig\ can be measured from SDSS aperture photometry without the need for delicate bulge-disk decompositions, success in using \DelSig\ would open the SDSS sample out to $z = 0.07$ to further bulge studies.

A sample of nearly 1000 SDSS galaxies from \citet{Gadotti2009} with careful bulge-disk-bar decompositions and measured values of the established bulge-classification parameter \Delmue\ is used to validate \DelSig.   \DelSig\ is compared to \Delmue\ for the \citetalias{Gadotti2009} sample and also to a larger mass-limited sample of SDSS central galaxies in the range 10.0 < log\ \mass/\msun\ < 10.4.   Results are as follows:

\begin{itemize}   
  
\item \DelSig\ and \Delmue\ (from \citetalias{Gadotti2009}) measure similar aspects of bulge structure, and derived bulge types, both P-bulge and C-bulge, are broadly similar. \DelSig\ can be used to provide statistically useful measures of bulge types for SDSS galaxies out to $z = 0.07$.

\item According to either \DelSig\ or \Delmue, the main distinction between bulge types is central stellar density: pseudo-bulges (P-bulges) have low central densities, while classical bulges (C-bulges) have high central densities.  

\item Additional SDSS parameters are compared to \DelSig\ and \Delmue. Structural parameters (based on kinematics and mass-density profiles) show fairly linear log-log relations \vs\ \Delmue\ and \DelSig\ with only moderate scatter.  Stellar-population parameters in contrast show a highly non-linear ``elbow'' in which specific star-formation rate remains roughly flat with central density and then falls rapidly at the elbow, where galaxies begin to quench. Similar trends are seen for both central and global star-formation indicators. The elbow seen here is the same as the feature seen by \citet{Fang2013} for SDSS galaxies and by \citet{Barro2017,Mosleh2017,Lee2018} for CANDELS galaxies.

\item In the mass-limited central SDSS sample, P-bulges are found to be a homogeneous class, are universally star-forming, and occupy the low-density portion of the horizontal elbow arm.  C-bulges in contrast exhibit a wide range of star formation rates from quenched to highly active and occupy the elbow itself and the vertical branch above it.  New terminology is introduced to subdivide C-bulges according to star-formation rate:  C-SFBs are star-forming classical bulges in and near the elbow, while C-QBs are quenched classical bulges at the end of the vertical branch. Preliminary evidence is mentioned that the detailed distributions of galaxies along and off the elbow may also vary with mass and environment.

\item Classifying galaxies as P-bulges or as C-bulges has been difficult in the past because criteria sometimes disagree.  Results here suggest that a major reason is the elbow-shaped correlations between structural parameters and star-formation indices, which means that classifying objects by structure as opposed to stellar-population indices will disagree for many galaxies.

\item Bimodality in bulge types has been discussed in the bulge-classification literature. A major conclusion from the present work is that distributions in spectral and structural parameters are both bimodal, stemming from the existence of two distinct ``islands" that are clearly visible in plots of spectral \vs\ structural parameters.  Understanding the bimodality of bulges will thus require modeling structure and stellar populations simultaneously. 

\end{itemize}

The structural and stellar-population relationships in this paper and their variation from index to index suggest an unsuspected richness in the central properties of galaxies as their star-formation rates begin to fade.  Exploring these in future may shed light on how galaxies quench. Galaxies with IFU data from surveys such as the SDSS-IV MaNGA may be useful in providing spatially resolved star formation maps to further understand the evolution track. If evolution is really along the horizontal branch of the elbow and then upwards to quenching, what is the expected signature in star formation maps?

\section*{Acknowledgements}
We acknowledge financial support from NSF grants AST-0808133 and AST-1615730. YL acknowledges support from the China Scholarship Council. This work is also partly supported by the National Key Basic Research and Development Program of China (No. 2018YFA0404501) and by the National Science Foundation of China (Grant No. 11821303, 11333003, 11390372 and 11761131004 for SM and YL). ARP acknowledges partial support from UNAM PAPIIT grant IA104118, the CONACyT  `Ciencia Basica' grant 285721, the UC-MEXUS postdoctoral fellowship and UC-MEXUS Collaborative Research Gran CN-17-125. 

Funding for the SDSS and SDSS-II has been provided by the Alfred P. Sloan Foundation, the Participating Institutions, the National Science Foundation, the U.S. Department of Energy, the National Aeronautics and Space Administration, the Japanese Monbukagakusho, the Max Planck Society, and the Higher Education Funding Council for England. The SDSS Web Site is http://www.sdss.org/.

GALEX (Galaxy Evolution Explorer) was a NASA Small Explorer, launched in 2003 April. We gratefully acknowledge NASA's support for construction, operation, and science analysis for the GALEX mission, developed in cooperation with the Centre National d'\'Etudes Spatiales of France and the Korean Ministry of Science and Technology.

We thank the referee for a thorough and stimulating referee report, which greatly improved the presentation. We also acknowledge our debt to Dmitri Gadotti for publishing his catalog of $\mu_e$ and other bulge-disk decomposition parameters, which provided the calibration sample needed for understanding the larger SDSS sample.

\bibliographystyle{mnras}
\bibliography{pseudobulge}

\appendix

\section{Reddening Corrections}
\label{sec:appendix}

\begin{figure}
    \includegraphics[width=\columnwidth]{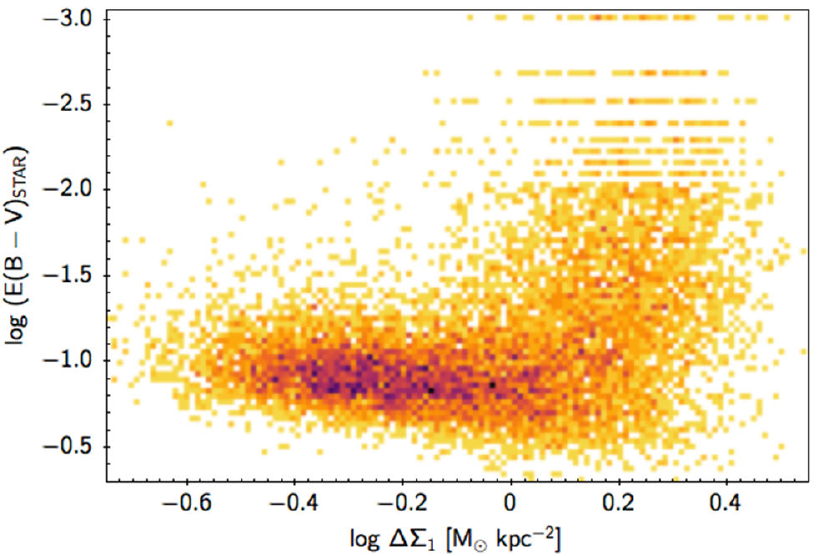}
    \caption{Reddening values for SDSS DR7 galaxies from  \citet{Oh2011} plotted \vs\ the bulge density parameter \DelSig. The sample used is the mass-limited SDSS sample in the mass range 10.0 < log\ \mass/\msun\ < 10.4 used in Figures \ref{fig:figure10}--\ref{fig:figure14}.  Reddening values are global and are estimated for the stellar continuum.  The Y-axis has been reversed to match the sense of Figures~\ref{fig:figure10}--\ref{fig:figure14}. The elbow pattern seen here matches similar patterns in Figures~\ref{fig:figure12}--\ref{fig:figure14}: the horizontal arm is populated by star-forming galaxies, and the vertical branch is populated by fading and quenched galaxies.  Reddening is moderate overall and does not vary greatly along the star-forming branch across the bottom; average values range from only 0.10 mag to 0.17 mag. Conclusions from Figures~\ref{fig:figure12}--\ref{fig:figure14} should therefore not depend sensitively on whether reddening corrections are applied or not.}
    \label{fig:figureA1}
\end{figure}

A concern is how reddening corrections may have affected the distributions of galaxies in Figures~\ref{fig:figure10}--\ref{fig:figure14}.  The reddening corrections of \citet{Oh2011} have been adopted here for colors (but not emission lines).  Figure~\ref{fig:figureA1} plots their $E(B-V)$ \vs\ \DelSig. These are global estimates for the stellar continua only. The mass range is limited to 10.0 < log\ \mass/\msun\ < 10.4, and the Y-axis is inverted to match the sense of Figures~\ref{fig:figure10}--\ref{fig:figure14}.  The striking elbow-shaped pattern in Figure~\ref{fig:figureA1} matches the similar pattern in Figures~\ref{fig:figure12}--\ref{fig:figure14}, which also plots spectral parameters.  The horizontal bottom branch contains star-forming galaxies, and the vertical branch is populated by fading and quenched galaxies, which in this figure have less or zero reddening.  

The important features of the figure are the overall small values of $E(B-V)$ and the small slope of the trend along the horizontal star-forming branch: values of $E(B-V)$ vary from 0.10 mag on average at low \DelSig\ to about 0.17 mag on average at high \DelSig.  Thus, $E(B-V)$ is small on average and rather constant, even for star-forming galaxies.  In the main text, we are interested in how star-forming indicators vary \vs\ \DelSig\ along this lower branch.  The conclusion from Figure~\ref{fig:figureA1}  is that our results should not depend greatly on whether raw or reddening-corrected parameters are used, and this was confirmed by separate tests.  Thus, we elect to use the reddening-corrected values of \citet{Oh2011} for all magnitudes and colors but use uncorrected values from SDSS DR7 for emission lines and spectral indices.

\section{Supplementary data}
\label{sec:data}

The \Sig\ value for SDSS DR7 galaxies with their SDSS identification can be found in the online version of this article. Notice that we only provide the \Sig\ value for galaxies with $0.02 < z < 0.07$ and axis-ratio \textit{b/a} > 0.5 to avoid the seeing degradation and the added opacity in edge-on galaxies.

\begin{table}
 \caption{Sample entries from the \Sig\ catalog for SDSS DR7 galaxies. The full table is available online as supplementary data.}
 \label{table:data}
 \begin{tabular}{|p{1.5cm}|p{1.5cm}|p{2.8cm}||p{0.8cm}|}
  \hline
  RA & DEC & SDSS DR7 objid & log \Sig \\
  \hline
  147.3295 & 0.0289 & 587725074995609782 & 9.4297 \\
  146.5128 & -0.8458 & 588848898824274111 & 9.4704 \\
  146.8644 &  -0.4641 &  588848899361276123 &  9.0138  \\ 
  146.7559 &  -0.1682 & 588848899898147178  & 8.2839 \\ 
  146.0937 &  -0.7931 &  587725073921343655 & 9.7409 \\ 
  \hline
  \end{tabular}
\small The first three columns show the SDSS identification of the galaxies. Column (4) shows the log \Sig\ value.\\
\end{table}

\bsp	
\label{lastpage}
\end{document}